\documentclass[twocolumn]{aastex63}
\usepackage{bm}
\usepackage{amsmath}

\RequirePackage{color}

\shorttitle{The global structure of the Milky Way's stellar halo}
\shortauthors{Sato \& Chiba}

\graphicspath{{./}{figures/}}

\begin{document}

\title{The global structure of the Milky Way's stellar halo based on the orbits of local metal-poor stars}

\correspondingauthor{Genta Sato}
\email{g.sato@astr.tohoku.ac.jp}

\author{Genta Sato}
\affiliation{Astronomical Institute, Tohoku University, Aoba-ku, Sendai 980-8578, Japan}

\author{Masashi~Chiba}
\affiliation{Astronomical Institute, Tohoku University, Aoba-ku, Sendai 980-8578, Japan}

\begin{abstract}
We analyze the global structure of the Milky Way (MW)'s stellar halo including its dominant subcomponent, Gaia-Sausage-Enceladus (GSE).
The method to reconstruct the global distribution of this old stellar component is to employ the superposition of the orbits covering over the large MW's space, 
where each of the orbit-weighting factor is assigned following the probability that the star is located at its currently observed position.
The selected local, metal-poor sample with ${\rm [Fe/H]}<-1$ using {\it Gaia} EDR3 and SDSS DR16 shows that the global shape of the halo is systematically rounder at all radii in more metal-poor ranges, such that an axial ratio, $q$, is nearly 1 for ${\rm [Fe/H]}<-2.2$ and $\sim 0.7$ for $-1.4<{\rm [Fe/H]}<-1.0$.
It is also found that a halo in relatively metal-rich range of ${\rm [Fe/H]}>-1.8$ actually shows a boxy/peanut-like shape, suggesting a major merger event.
The distribution of azimuthal velocities shows a disk-like flattened structure at $-1.4<{\rm [Fe/H]}<-1.0$, which is thought to be the metal-weak thick disk.
For the subsample of stars showing GSE-like kinematics and at ${\rm [Fe/H]}>-1.8$,
its global density distribution is more spherical with $q \sim 0.9$ than the general halo sample, having an outer
ridge at $r\sim20$~kpc.
This spherical shape is consistent with the feature of accreted halo components
and the ridge suggests that the orbit of GSE's progenitor has an apocenter of $\sim 20$~kpc. 
Implications for the formation of the stellar halo are also presented.
\end{abstract}


\section{Introduction} \label{sec:intro}
It is generally thought that the stellar halo of the Milky Way (MW) contains the fossil record on the MW's formation history imprinted in
its kinematical and chemical properties.
Since \citet{eggen1962} showed a model of the MW's formation process from the kinematic analysis of halo stars in the solar neighborhood,
many studies of this fossil Galactic component have been carried out using a large number of newly available halo sample, based on considerable observational efforts for their assembly over the past decades.
These include, for example, the Hipparcos Catalog obtained from the first astrometry satellite \citep{perryman1997}, and spectroscopic catalogs such as
RAVE \citep{steinmetz2006}, SEGUE \citep{yanny2009}, LAMOST \citep{cui2012, zhao2012}, GALAH \citep{desilva2015}, APOGEE \citep{majewski2017},
H3 \citep{conroy2019}, and more.
Perhaps the most significant impacts on this field of research have been brought by the second astrometry satellite, {\it Gaia}.
{\it Gaia} catalog \citep{gaiacollabo2016, gaiacollabo2018, gaiacollabo2021} provides trigonometric parallaxes and proper motions for billions of Galactic stars with unprecedented high accuracy.
Based on these astrometry data of stars in {\it Gaia} catalog combined with spectroscopic catalogs, the MW's new dynamical maps have been drawn \citep[e.g.][]{helmi2018,belokurov2018, myeong2018, myeong2018b, beane2019, hagen2019, iorio2019, anguiano2020, cordoni2021, koppelman2021},
and new aspects of the MW's formation and evolution history have been revealed \citep[e.g.][]{antoja2018, sestito2019, wyse2019}.
In particular, {\it Gaia} catalog has enabled to discover new substructures in the MW's stellar halo \citep[e.g.][]{koppelman2019, myeong2019, li2020},
which are remnants of past merging/accretion events associated in the MW's formation history \citep[e.g.][]{fernandeztrincado2020, naidu2020, naidu2021}.

These observational information of the MW's stellar halo have been investigated in comparison with various numerical simulations,
which produce a number of theoretically predicted halo structures in the MW-sized galaxies,
including their global density profiles and dynamical motions over a large halo space \citep[e.g.][]{rodriguez2016, liang2021, monachesti2019}
as well as the merging/accretion history of satellite galaxies \citep[e.g.][]{cooper2010, mackereth2019, kruijssen2020, santistevan2020}.

One of the most significant substructures in the MW's stellar halo revealed in {\it Gaia} catalog is Gaia-Sausage-Enceladus (GSE) \citep{helmi2018, belokurov2018}.
GSE shows an elongated distribution in a velocity space with azimuthal velocities of $v_\phi\sim0$
and is thought as a tracer of a merger event with a SMC-class galaxy during the MW's evolution about 10~Gyr ago \citep{helmi2018}.
The further details of GSE's dynamical and chemical properties have been determined observationally \citep[e.g.][]{feuillet2020, monty2020},
and several simulations have also been explored to constrain the properties of GSE's progenitor \citep[e.g.][]{bignone2019, koppelman2020, kim2021}.
\citet{evans2020} showed that a merger event with a GSE-like progenitor is rare for MW-like galaxies.
Thus, GSE is a unique feature of the MW's stellar halo, 
thereby suggesting that the merger should be essential for the MW's evolution history.

In previous observational studies, a stellar sample selected and used in the dynamical analysis is generally confined in the local volume of the Galactic space,
so the global structure of the stellar halo is hardly derived.
For example, the region where {\it Gaia} catalog obtains the accurate parallax data is limited near the Sun, within heliocentric distances of $\sim4$~kpc (see Subsection \ref{subsec:data}).
To solve this issue and derive the spatial distribution of halo stars over a large Galactic space,
we adopt the method based on stellar orbits \citep{may1986, SLZ1990, chiba2000, chiba2001}.
This method is designed, based on the superposition of stellar orbits, for reconstructing the stellar distribution over a large region,
up to the apocentric distances where stars can reach.
\citet{chiba2000} used about 1200 metal-poor stars mostly selected from Hipparcos catalog for this method. To step further in this work, 
we adopt {\it Gaia} Early Data Release 3rd (EDR3) \citep{gaiacollabo2021}
in combination with the spectroscopic information from SDSS DR16 SEGUE, which allows us to have a much larger number of stars with better astrometric accuracy.
Thus, it is expected that our reconstruction of the global halo distribution, over $r \sim 30$~kpc, is more accurate and statistically more significant than previous studies.

Moreover, we also apply this method to the stars showing GSE-like kinematics
and derive GSE's global structure in the Galactic halo. 
In contrast to previous works \citep[e.g.][]{feuillet2020, monty2020}, which investigated GSE's properties based on their local observational information, we adopt the strategy using the superposition of orbits for GSE-like member stars to construct the GSE's global distribution for the first time.

This paper is organized as follows.
Section \ref{sec:meth} explains the observational data with the selection conditions we adopt and the method to reconstruct the distribution of the MW's stellar halo as well as the GSE-like subsample.
In Section \ref{sec:rslt}, we show the structure of the reproduced global halo distribution and the results of the fitting to an ellipsoidal profile characterized by an index of the power-law radial profile, $\alpha$, and an axial ratio, $q$.
In Section \ref{sec:dscs}, we discuss the implication from the derived distribution and structures of the stellar halo, and infer, based on the current results, about the MW's formation and evolution history.
Section \ref{sec:cncl} summarizes our work and conclusions.

\section{Method} \label{sec:meth}
\subsection{Data} \label{subsec:data}

We adopt {\it Gaia} EDR3 Catalog \citep{gaiacollabo2021} for astrometry data
and SDSS DR16 SEGUE catalog \citep{juric2008, ivezic2008, bond2010} for spectroscopic data.
We extract the common stellar data from these catalogs satisfying the following five conditions
to adopt for our analysis.

The first condition is that the stellar metallicity ${\rm [Fe/H]}$ is more metal-poor than $-1$.
This condition is to select stars belonging to the fiducial stellar halo having low metals and possibly old ages.

The second and third conditions are that 
the effective temperature and the surface gravity of a star are within the range $4,500 < T_{\rm eff} < 7,000$~K
and $3.0 < \log{g} < 5.2$, respectively.
These constraints are to extract the main sequence stars, following the work of \citet{ivezic2008},
which also derived the range of the observational quantities that fulfil main sequence stars.

The fourth and fifth conditions are for ensuring the accuracy of the data, i.e.,
to avoid stars having large uncertainties in parallaxes and metallicities, respectively.
We select the stars having the relative uncertainty of the parallax $\sigma_\varpi / \varpi$ smaller than 0.2
and the uncertainty of ${\rm [Fe/H]}$ smaller than 0.1~dex.

Next, we carry out cross-match between these two catalogs using the CDS Xmatch Service
\footnote{\url{http://cdsxmatch.u-strasbg.fr}},
where the cross-matching condition for the stellar position is given to be smaller than 1~arcsec.
Setting only this positional condition however leaves some cases that
a few number of SDSS stars match with a single {\it Gaia} star.  
In this case, we impose an additional condition for the photometry of stars, following
the relation between the $G$ band used in {\it Gaia} and $(g, i)$ bands in SDSS \citep{busso2021}:

\begin{equation}
    \begin{split}
G-g = &-0.1064 -0.4964(g-i)  \\
      &-0.09339(g-i)^2 +0.004444(g-i)^3
    \end{split}
\label{eq:band_Gg}
\end{equation}
We adopt the star having smallest magnitude difference among the position-matched candidate stars between the two catalogs.

\begin{figure}[t]
\centering
\includegraphics[width=62mm]{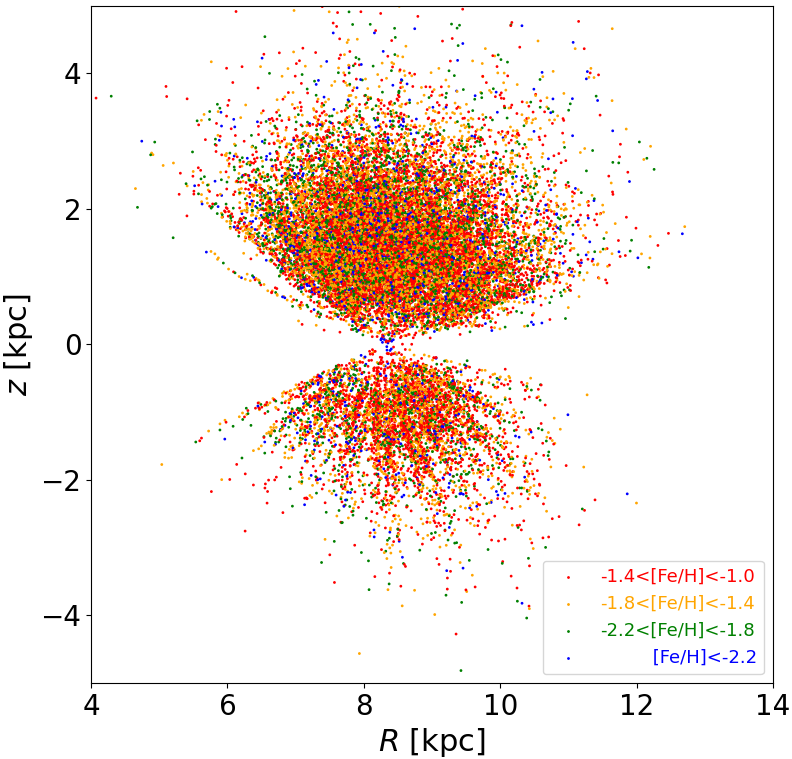}
\includegraphics[width=60mm]{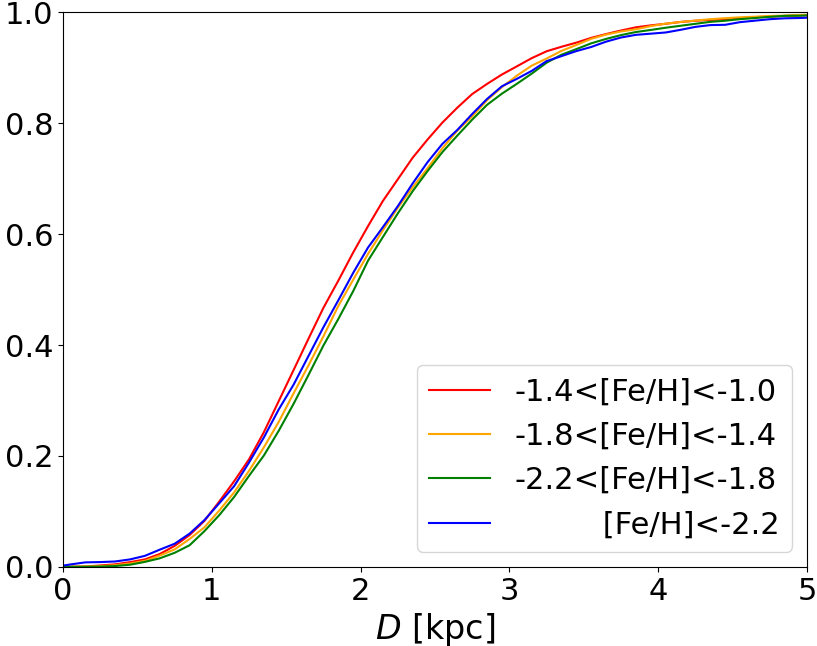}
\caption{Upper panel: The meridian distribution of fiducial metal-poor halo stars adopted in this work selected from the cross-match of {\it Gaia} EDR3 with SDSS DR16 SEGUE catalogs.
Different colors correspond to different ranges of stellar metallicities:
red for $-1.4<{\rm [Fe/H]}<-1.0$, orange for $-1.8<{\rm [Fe/H]}<-1.4$, green for $-2.2<{\rm [Fe/H]}<-1.8$, and blue for ${\rm [Fe/H]}<-2.2$.
This correspondence is applied in following figures.
Bottom panel: The cumulative number distribution of the heliocentric distance, $D$, for the adopted stellar sample divided by the metallicity ranges. 
The histograms are normalized so that the total number is 1 for each range. 
\label{fig:Rz}}
\end{figure}

This spectroscopic catalog (SDSS SEGUE) is used to extract line-of-sight velocities and spectroscopic
metallicities of sample stars. Stars containing these spectroscopic information with good accuracy are limited
to only bright ones, so the range of $G$-band absolute magnitude of our sample is $M_G\lesssim8$~mag.

We have extracted 25,093 stars as the fiducial members of the stellar halo
\footnote{In comparison, \citet{kim2021-2} recently adopt $\sim$~500,000 thick-disk and halo-like stars
with $M_G<13$~mag to analyze their reduced proper motions without the limitation of the availability of
spectroscopic data.}.
The distributions of these stars are shown in Figure \ref{fig:Rz}.
In the following, we divide the sample stars mainly into four subsamples based on their metallicities,
$-1.4<{\rm [Fe/H]}<-1.0$, $-1.8<{\rm [Fe/H]}<-1.4$, $-2.2<{\rm [Fe/H]}<-1.8$, and ${\rm [Fe/H]}<-2.2$
(shown with different colors in Figure \ref{fig:Rz} and following figures).
The number of stars in each subsample is 10,214, 8,484, 4,518, and 1,877, respectively.

Additionally, we also consider the likely member stars belonging to the Gaia-Sausage-Enceladus (GSE).
The GSE stars are distributed around the azimutal velocities, $v_\phi$, near 0 \citep{belokurov2018, myeong2018},
which corresponds to the vertical angular momentum $L_z$ as small as 0.
We note that strongly-bound stars, namely stars with low orbital energy, are dominated by the so-called {\it in-situ} halo component \citep[e.g.][]{naidu2020},
and such stars are not GSE's members.
Thus, we extract the GSE-like stars, which fulfil the conditions about the binding energy $E$ and the vertical angular momentum $L_z$;
$E>-150,000~{\rm km^2~s^{-2}}$ and $-500<L_z<500~{\rm kpc~km~s^{-1}}$
(see Subsection \ref{subsec:pote} for the definition of $E$ and $L_z$).
The number of the extracted stars is 9,402.

These stars belonging to the GSE-like members are also divided into four subsamples depending on the metallicity ranges of
$-1.4<{\rm [Fe/H]}<-1.0$, $-1.8<{\rm [Fe/H]}<-1.4$, $-2.2<{\rm [Fe/H]}<-1.8$, and ${\rm [Fe/H]}<-2.2$,
where the number of stars are 3,269, 3,607, 1,827, and 699, respectively.
Although the real GSE may dominate in the intermediate metallicity range, e.g., $-1.66<{\rm [Fe/H]}<-1.33$ \citep{belokurov2018}, covering our subsample of $-1.8<{\rm [Fe/H]}<-1.4$, 
we also consider other metallicity ranges to analyze the metallicity dependence of the GSE-like structure and to compare with the general distribution of the stellar halo.

\subsection{Dynamical Model} \label{subsec:model}
We adopt the method based on the superposition of the discrete stellar orbits to reproduce and analyze the global distribution of the MW's stellar halo following the seminal works by \citet{may1986} and \citet{SLZ1990}.
Although {\it Gaia} catalog provides billions of astrometric data,
stars with accurate parallaxes of $\sigma_\varpi / \varpi < 0.2$ are available only in the local volume near the Sun, 
which is small compared with the entire MW's halo space (Figure~\ref{fig:Rz}).
By calculating the stellar orbits and their superposition with appropriate weighting, 
we can construct the MW's halo distribution beyond the region where we can observe stars locally and individually.
This subsection presents the adopted gravitational potential, the properties of stellar orbits in it, and then
the method to construct the global distribution of the stellar halo from the orbits.

\subsubsection{Gravitational potential}\label{subsec:pote}
In this work, for the purpose of calculating each orbital density analytically as described below,
we adopt the St\"{a}ckel form for the MW's gravitational potential.
The St\"{a}ckel potential, $\psi$, is generally expressed as

\begin{equation}
\psi(\lambda, \nu)=-\frac{(\lambda+\gamma)G(\lambda)-(\nu+\gamma)G(\nu)}{\lambda-\nu}
\label{eq:Stackel}
\end{equation}
where $G(\tau)$ is an arbitrary function \citep{dezeeuw1985,dejonghe1988}.
$\lambda$ and $\nu$ are the spheroidal coordinates system ($\lambda, \nu$),
defined as the solution to the following equation for $\tau$,

\begin{equation}
\frac{R^2}{\tau +\alpha}+\frac{z^2}{\tau +\gamma}=1
\label{eq:def_ln}
\end{equation}
where $\alpha$ and $\gamma$ are constants and $(R, z)$ are the Galactocentric cylindrical coordinates,
assuming that the position of the Sun is $(R_\odot, z_\odot)\equiv(8.3$~kpc, 0) \citep{gillessen2009}.
The contours of $\lambda, \nu =$
constant describe ellipses and hyperbolas, respectively.

\begin{figure}[t]
\centering
\includegraphics[width=80mm]{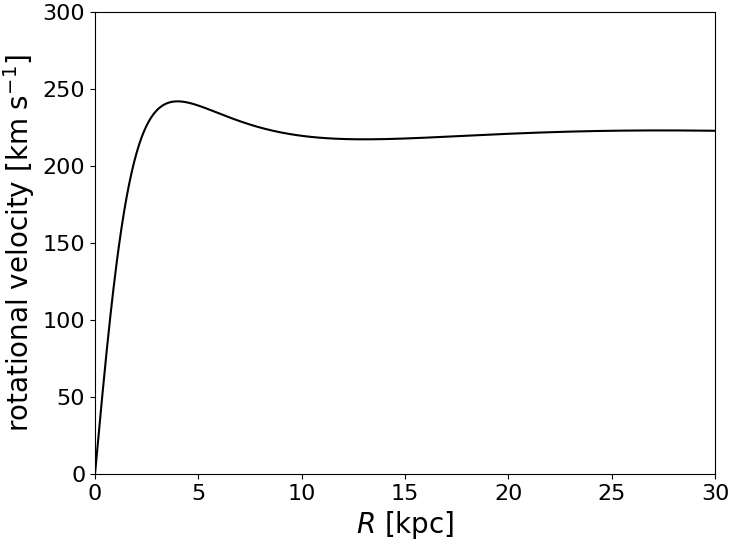}
\caption{The rotation curve of the mass model adopted in this work, that is basically in agreement with
the observed one in the MW.
\label{fig:rotcurve}}
\end{figure}

For the arbitrary function $G(\tau)$ in Equation \eqref{eq:Stackel}, 
we consider the contribution of gravitational force from the stellar disk and dark halo components. 
Here we adopt a Kuzmin-Kutuzov potential
\citep{dejonghe1988, batsleer1994, chiba2001} for both components given in the first and second terms below, respectively,

\begin{equation}
G(\tau)=\frac{G_{\rm grav}kM}{\sqrt{\tau}+\sqrt{-\gamma}}+\frac{G_{\rm grav}(1-k)M}{\sqrt{\tau+b}+\sqrt{-\gamma+b}}
\label{eq:Gtau}
\end{equation}
where $G_{\rm grav}$ is the gravitational constant, $M$ is the total mass of the MW, $k$ is the ratio of the disk mass to the total mass, and $b$ is the parameter to make this two-component model in the St\"{a}ckel form.

In this work, we adopt $M=10^{12}M_\odot$, which is nearly in agreement with several recent measurements, e.g.,
$0.7 \times 10^{12}M_\odot$ \citep{eadie2019}, $1.3 \times 10^{12}M_\odot$ \citep{posti2019, grand2019}, $1.0 \times 10^{12}M_\odot$ \citep{deason2019}, and $1.2 \times 10^{12}M_\odot$ \citep{deason2021}.
For other parameters, $k$, $\alpha$, $\gamma$, and $b$, we take the same values as those in \citet{chiba2001}
so as to make the current model resemble the real Galaxy: 
the circular velocity at $R>4$~kpc is constant and $\simeq 220$~km~s$^{-1}$, the local mass density at $R=R_{\odot}$ is $0.1 \simeq 0.2$ $M_{\odot}$~pc$^{-3}$,
and the surface mass density at $R=R_{\odot}$ is 70 $M_{\odot}$~pc$^{-2}$ within $z \lesssim 1.1$~kpc. We adopt
$k=0.09$ for the disk fraction, and $\alpha$ and $\gamma$ are associated with the disk's scalelength in $R$, $a_D$,
and scaleheight in $z$, $c_D$, respectively, as $\alpha = -a_D^2$ and $\gamma = -c_D^2$. 
The corresponding scales for the dark halo, $a_H$ and $c_H$, are set as $a_H^2 = a_D^2 + b$ and $c_H^2 = c_D^2 + b$.
These values of scales are given as $c_D=0.052$~kpc, $c_H=17.5$~kpc, $c_D/a_D=0.02$, and $c_H/a_H=0.99$.

\begin{figure}[t]
\centering
\includegraphics[width=80mm]{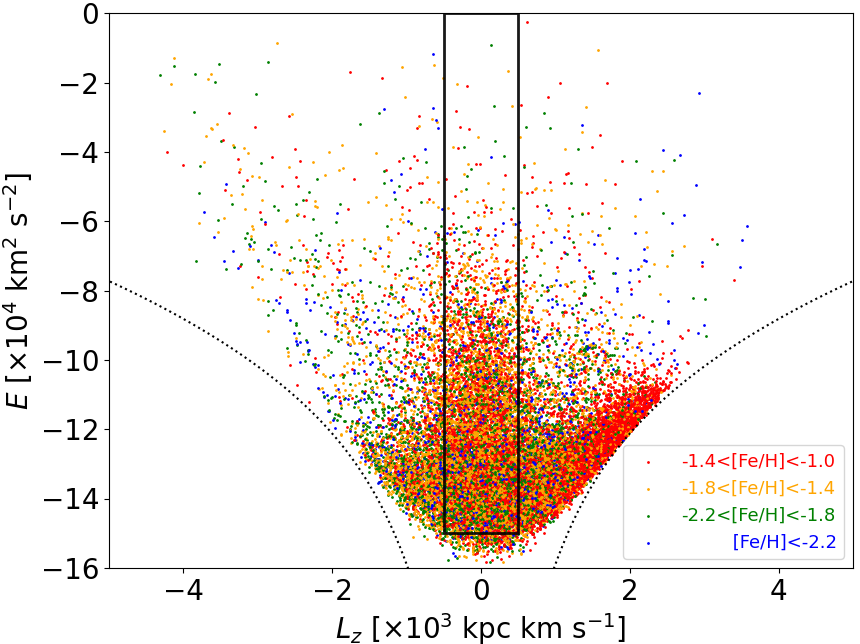}
\caption{The phase-space $(E, L_z)$ distribution of fiducial halo stars adopted in this work.
The black dotted line corresponds to a circular orbit, i.e., the lowest orbital energy, $E$, for
a given vertical component of angular momentum, $L_z$.
The rectangle indicates the selection region of stars we adopted as GSE-like subsample.
\label{fig:ELz_sau}}
\end{figure}

The rotation curve of this gravitational potential is shown in Figure \ref{fig:rotcurve}.
It captures the feature that the rotational speed near the Sun, $R \sim 8$~kpc, is about 220~km~s$^{-1}$
and the curve farther than 8~kpc is almost flat.
They are consistent with the observational properties \citep[e.g.][]{mroz2019, sofue2020},
so we believe this model is appropriate for the following analysis.
It is true that the current St\"{a}ckel potential does not perfectly represent the actual Galactic potential, especially near the disk plane, in contrast to other models such as MWPotential2014 \citep{bovy2015}.
However, since we focus on the distribution of the stellar halo, which is mostly spread out in high-$z$ region widely,
we consider that the difference of the potentials near the disk does not affect the result critically.

In this work, we define the heliocentric distance as the inverse of parallax.
\citet{bailer2021} showed the more accurate distance considering the effect of extinction, magnitude and color.
We find that the difference between $\varpi^{-1}$ and the distance of \citet{bailer2021} lies within 1$\sigma$ for 75\% of our sample and within 2$\sigma$ for 90\%,
under the condition of $\sigma_\varpi/\varpi<$~0.2.
Thus we regard the inverse of parallax as the distance in this work.

\subsubsection{Stellar orbits}\label{subsec:orbit}
\begin{figure}[t]
\centering
\includegraphics[width=80mm]{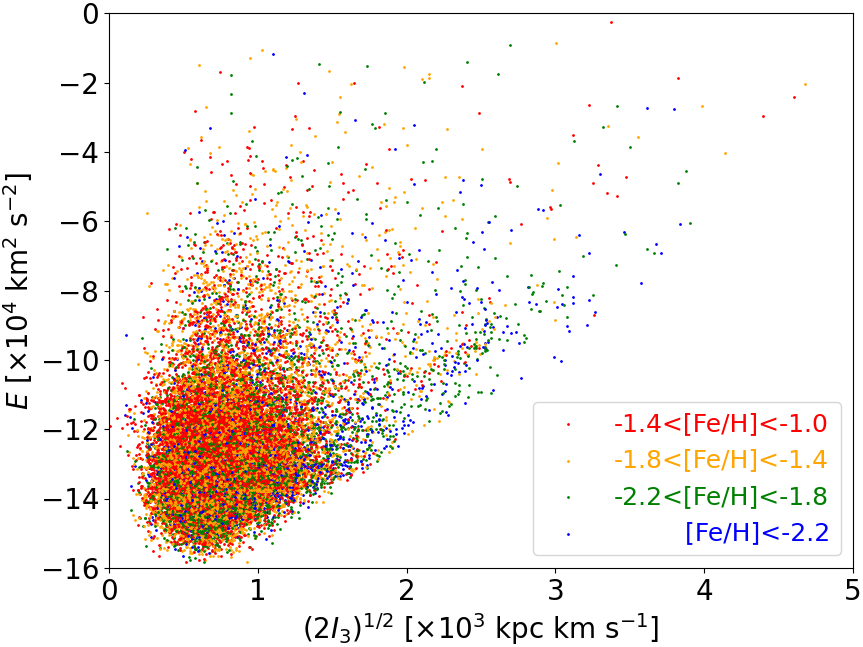}
\caption{The distribution of fiducial halo stars in the phase-space $(\sqrt{2I_3}, E)$.
We note that $\sqrt{2I_3}$ has the dimension of angular momentum and becomes $\sqrt{L_x^2+L_y^2}$ in the case of a spherically symmetric gravitational potential.
\label{fig:EI3}}
\end{figure}

The St\"{a}ckel potential has a unique feature that 
all three integrals of motion can be written analytically \citep{dezeeuw1985b}.
This axisymmetric system allows three integrals of motion, $(E, I_2, I_3)$.
$E$ and $I_2$ are defined as

\begin{align}
E & = \frac{1}{2}|{\bm v}|^2 +\psi({\bm x})
\label{eq:ene}\\
I_2 & = \frac{1}{2}L_z^2~{\rm with} ~ L_z = R v_\phi
\label{eq:zAngMome}
\end{align}
where $v_\phi$ is the azimuthal component of the velocity vector.
The remaining third integral cannot generally be described analytically in any axisymmetric system,
except for the case of a St\"{a}ckel potential.
In the case of a St\"{a}ckel model, the third integral $I_3$ is expressed as

\begin{equation}
I_3=\frac{1}{2}(L_x^2+L_y^2)+(\gamma-\alpha)\left[\frac{1}{2}v_z^2-z^2\frac{G(\lambda)-G(\nu)}{\lambda-\nu}\right]
\label{eq:I3}
\end{equation} 
where $(x, y, z)$ are the Cartesian coordinates with $x=R\cos\phi$ and $y=R\sin\phi$.

The distribution of our stellar data in phase space $(E, I_2, I_3)$ is shown in Figure \ref{fig:ELz_sau} and \ref{fig:EI3}.
The rectangle in Figure \ref{fig:ELz_sau} means that stars inside it belong to the GSE-like subsample.

The physical meaning of $I_3$ is the generalized horizontal angular momentum $L_\parallel$:
in a spherically symmetric case, two constants $\alpha$ and $\gamma$ are equal, so $I_3 =(1/2)(L_x^2+L_y^2)=
(1/2)L^2_\parallel=(1/2)(L^2-L_z^2)$. 
The total angular momentum $L$ is an integral of motion in a spherical model,
so $L_\parallel$ is also an integral of motion.
Therefore, $\sqrt{2I_3}$ can be regarded as the generality of $L_\parallel$
(This is why the horizontal axis in Figure \ref{fig:EI3} describes $\sqrt{2I_3}$, not simply $I_3$).

\begin{figure}[t]
\centering
\includegraphics[width=80mm]{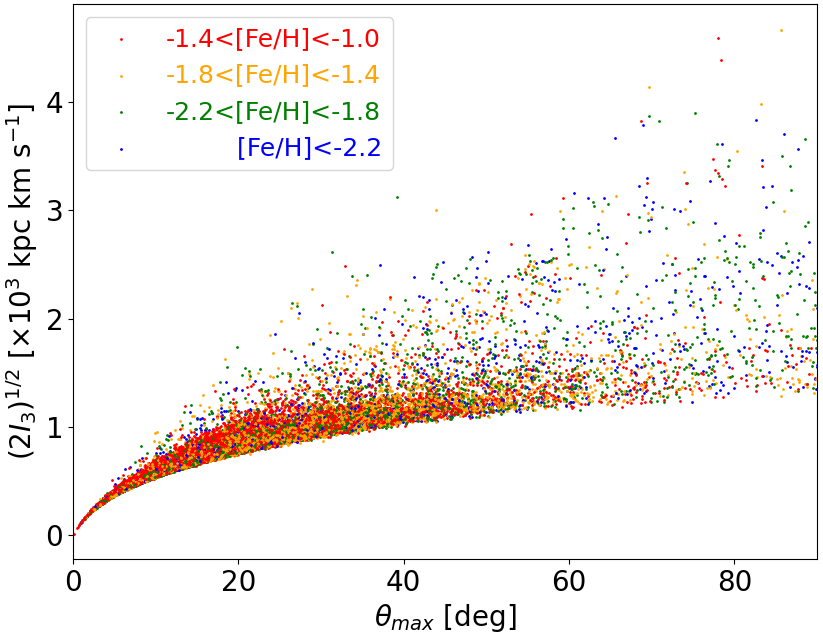}
\caption{The correlation between orbital angles measured from the Galactic plane, $\theta_{\rm max}$, and $\sqrt{I_3}$ in fiducial halo stars.
\label{fig:thI3}}
\end{figure}

\begin{figure}[t]
\centering
\includegraphics[width=80mm]{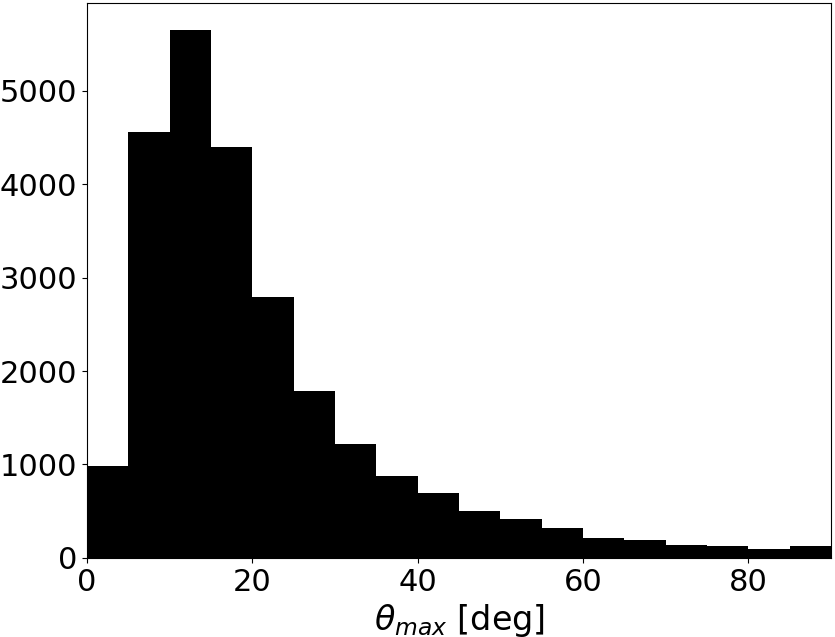}
\caption{The distribution of orbital angle, $\theta_{\rm max}$.
While the number of stars with $\theta_{\rm max}>15^\circ$ shows a monotonous decrease with increasing $\theta_{\rm max}$,
the stars with $\theta_{\rm max}<15^\circ$ are deficient from the extrapolation from this distribution.
\label{fig:histth0}}
\end{figure}

It it known that $I_3$ has the correlation with the maximum inclination angle, $\theta_{\rm max}$, of an orbit,
where $\theta_{\rm max}$ is defined as the angle from the Galactic plane.
Our $\theta_{\rm max}$ vs. $I_3$ diagram is shown in Figure \ref{fig:thI3}.
We find that the correlation appears clearly.
Additionally, Figure \ref{fig:histth0} shows the number distribution of $\theta_{\rm max}$ for the current sample stars.
It suggests that the distribution is peaked at $\theta_{\rm max} \sim 15^\circ$ and smoothly declining at larger $\theta_{\rm max}$.

\begin{figure*}[t]
\centering
\includegraphics[width=80mm]{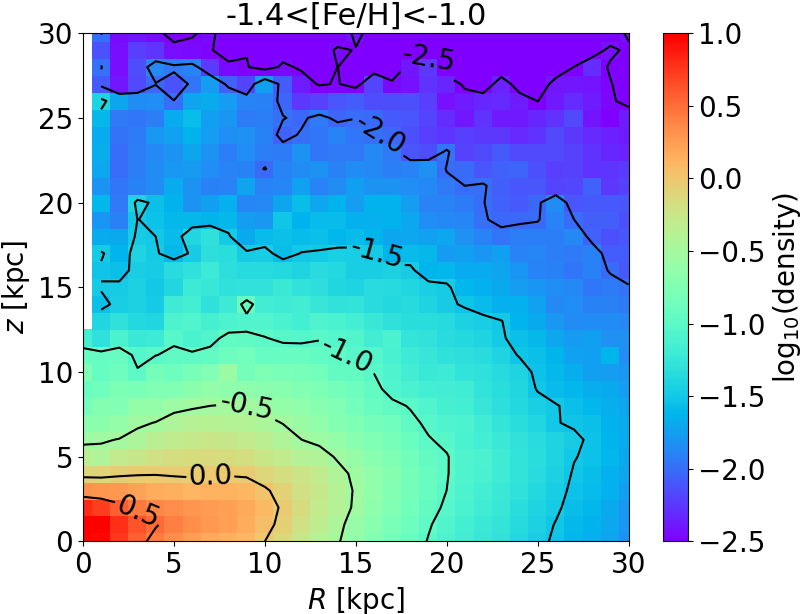} \includegraphics[width=80mm]{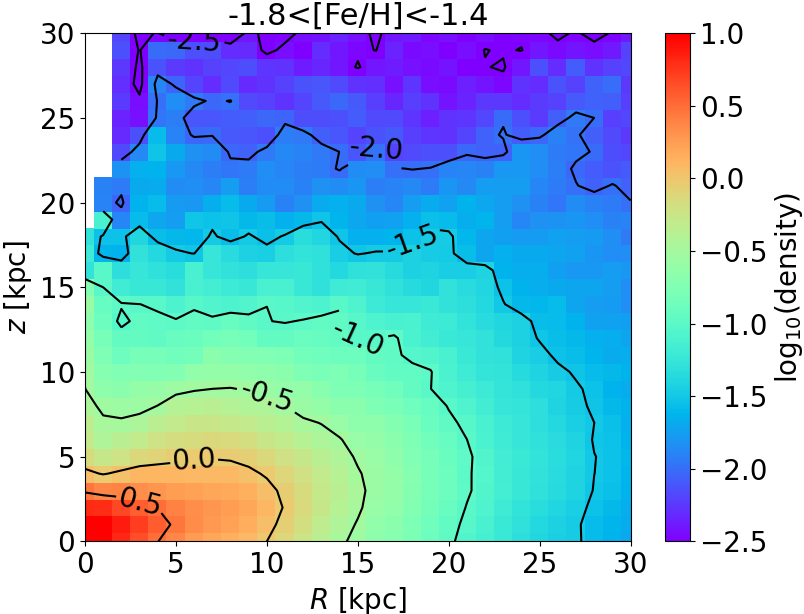}
\includegraphics[width=80mm]{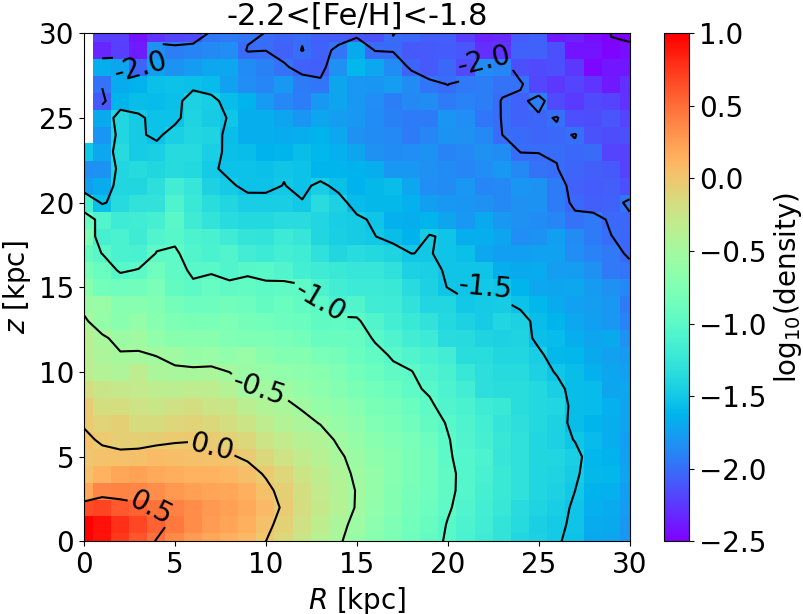} \includegraphics[width=80mm]{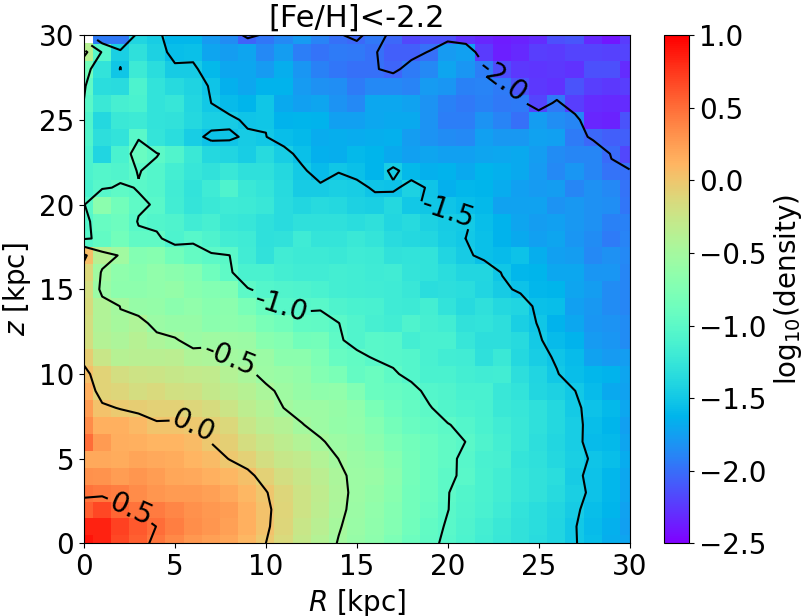}
\caption{The constructed global density distribution calculated from the entire halo sample for the four different metallicity ranges,
$-1.4<{\rm [Fe/H]}<-1.0$ (upper left), $-1.8<{\rm [Fe/H]}<-1.4$ (upper right),
$-2.2<{\rm [Fe/H]}<-1.8$ (lower left), and ${\rm [Fe/H]}<-2.2$ (lower right panel).
The distribution is normalized at $(R, z)=(10~{\rm kpc}, 0)$.
\label{fig:gene_coar}}
\end{figure*}

Since the integrals of motion contain complete information about the orbit of a star,
the fact that all integrals of motion are described analytically means that we can reproduce the stellar orbit accurately and completely. 
Under the St\"{a}ckel potential, 
the orbital density of $i$-th star at position ${\bm x}$ is expressed as \citep{dezeeuw1985, dejonghe1988, SLZ1990, chiba2001}

\begin{align}
\rho_{orb,i}&({\bm x})=\frac{2\sqrt{2}}{R}\frac{1}{\sqrt{N_i(\lambda)(\lambda+\gamma)}\sqrt{-N_i(\nu)(\nu+\gamma)}\sqrt{I_{2,i}}}
\label{eq:rhoorb}\\
&{\rm where}~N_i(\tau)=G(\tau)-\frac{I_{2,i}}{\tau+\alpha}-\frac{I_{3,i}}{\tau+\gamma}+E_i
\label{eq:Ntau}
\end{align}
We reconstruct the density distribution of the sample halo stars, $\rho({\bm x})$,
by overlaying the orbital densities with proper weights, $c_i$.

\begin{equation}
\rho({\bm x})=\sum_{i=1}^N c_i \rho_{orb,i}({\bm x})
\label{eq:sumrho}
\end{equation}
Here, the orbit-weighting factors $c_i$ are determined by a maximum likelihood method as follows.
When the $i$-th star has integrals of motion ($E_i,~I_{2,i},~I_{3,i}$),
the probability $P_{ij}$ that the star is observed at the position ${\bm x}_j$ is expressed as

\begin{equation}
P_{ij}=\frac{c_i \rho_{orb,i}({\bm x}_j)}{\sum_{k=1}^N c_k \rho_{orb,k}({\bm x}_j)}
\label{eq:Pij}
\end{equation}
The $i$-th integrals of motion are calculated from the $i$-th observed star,
so the case of $i=j$ realizes for any $i$.
Therefore the probability $P_{ij}$ should be maximized at $i=j$ for arbitrary $j$.
We define the weights $c_i$ so that $\prod_{i=1}^N P_{ii}$ is maximized \citep{SLZ1990, chiba2000, chiba2001}.

The orbit weighting factor $c_i$ is designed to be proportional to the inverse of the time that the $i$-th star spends
at the currently observed position,
or the inverse of the $i$-th orbital density at the position where the $i$-th star is observed, $\rho_{orb,i}({\bm x}_i)$.

\begin{equation}
c_i\propto\frac{1}{\rho_{orb, i}({\bm x}_i)}
\label{eq:ci_prop}
\end{equation}

This procedure means that a higher weight is given to a star located at the position of lower $\rho_{orb,i}({\bm x}_i)$ in the
course of its orbital motion: since the probability to find such a star is low at its current position, the fact that it is
actually observed there leads to the assignment of a higher orbit-weighting factor $c_i$.
Conversely. a lower $c_i$ is assigned to a star currently located at higher $\rho_{orb,i}({\bm x}_i)$, e.g.,
near the edge of its orbit, where a star stays for a longer time.

In addition to the global density structure of metal-poor halo stars, we also construct their global rotational velocity
distribution.
We calculate the values of mean rotational velocity $\langle v_\phi \rangle$ as follows.

\begin{align}
\langle v_\phi&({\bm x}) \rangle = \frac{1}{\rho({\bm x})} \sum_i c_i \rho_{orb,i}({\bm x}) v_{\phi,i}({\bm x})
\label{eq:muvtau}\\
&{\rm where}~v_{\phi,i}({\bm x}) = \pm \sqrt{\frac{2I_{2,i}}{R^2}}
\label{eq:vphix}
\end{align}
The sign in Eq. (\ref{eq:vphix}) is determined to match the rotational direction of the observational $i$-th star.

\section{Results} \label{sec:rslt}
In what follows, we show our results on the meridian plane using the coordinates ($r, \theta$) and also the cylindrical coordinates ($R, z$).
The relation between these coordinates is $R=r\cos\theta, z=r\sin\theta$,
so $(r, \theta)$ correspond to the 2-dimensional polar coordinates 
where $\theta$ is measured from the Galactic plane.

\subsection{The entire halo sample} \label{subsec:whole}
\subsubsection{Density distribution} \label{subsec:whl_dens}
The reproduced global density distribution of the halo sample expressed by Eq.\eqref{eq:sumrho} is shown in Figure \ref{fig:gene_coar}, 
where the total amplitude of the density is normalized by the value at $(R, z)=(10~{\rm kpc}, 0)$.
The density contours hold nearly ellipsoidal shapes in all metallicity ranges, 
although the detailed contour pattern looks like a somewhat boxy or peanut-like shape.
In order to characterize these global density distributions quantitatively, 
we fit each of them to a simple ellipsoidal profile: 

\begin{equation}
\rho(\bm x)=\left[\frac{R_c^2}{R^2 + (z/q)^2}\right]^{\alpha /2}
\label{eq:sing_ellip}
\end{equation}
where $R_c(=10$~kpc$)$ denotes the radius of the density normalization.

\begin{figure}[t]
\centering
\includegraphics[width=80mm]{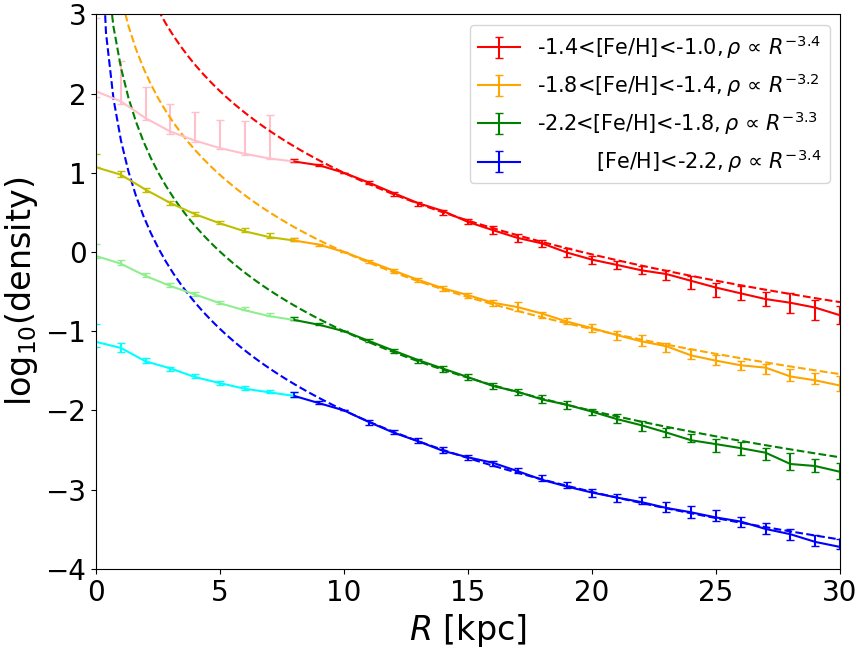}
\caption{The density distribution along the Galactic plane, extracted from Figure \ref{fig:gene_coar}.
The solid lines show the reproduced distribution and the dashed lines present the fitting curves with a best fitted power-law slope, $\alpha$.
The colors of lines correspond to the range of metallicity.
The light colors indicate the region at $R<8$~kpc, which are excluded in the fitting to the profile $\rho\propto R^{-\alpha}$.
Each line is shifted vertically for the purpose of comparing between different metallicity ranges.
The error bars show the uncertainties propagating from the observational errors associated with the position on the sky, parallax, proper motion and line-of-sight velocity.
The error bars in the following figures also include all of these observational errors.
\label{fig:gene_pla}}
\end{figure}

\begin{figure}[t]
\centering
\includegraphics[width=80mm]{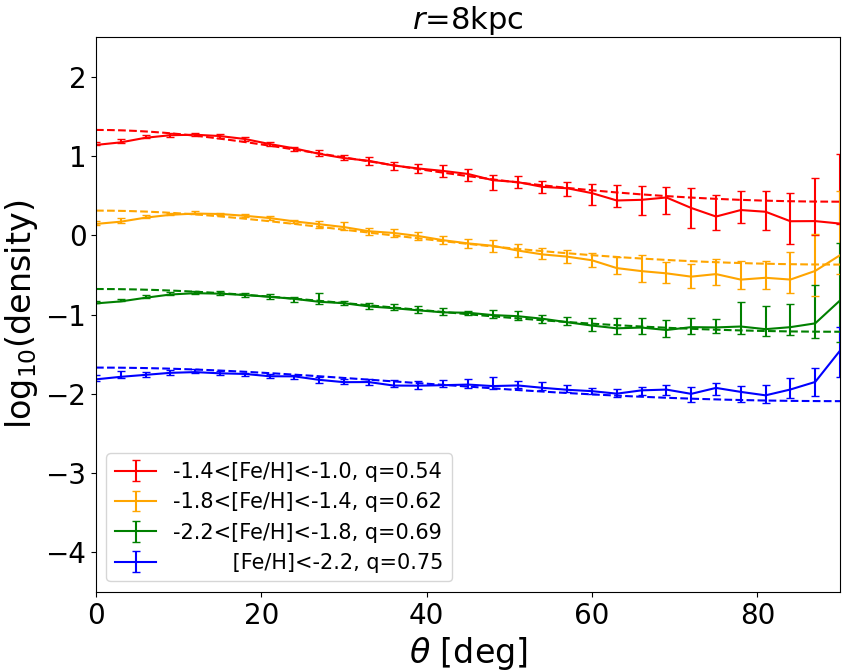}
\includegraphics[width=80mm]{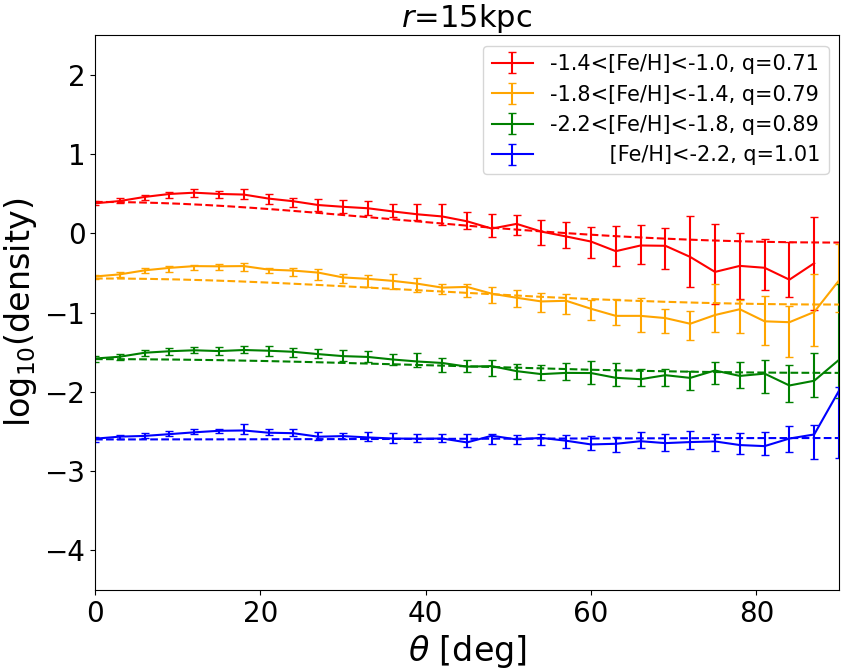}
\includegraphics[width=80mm]{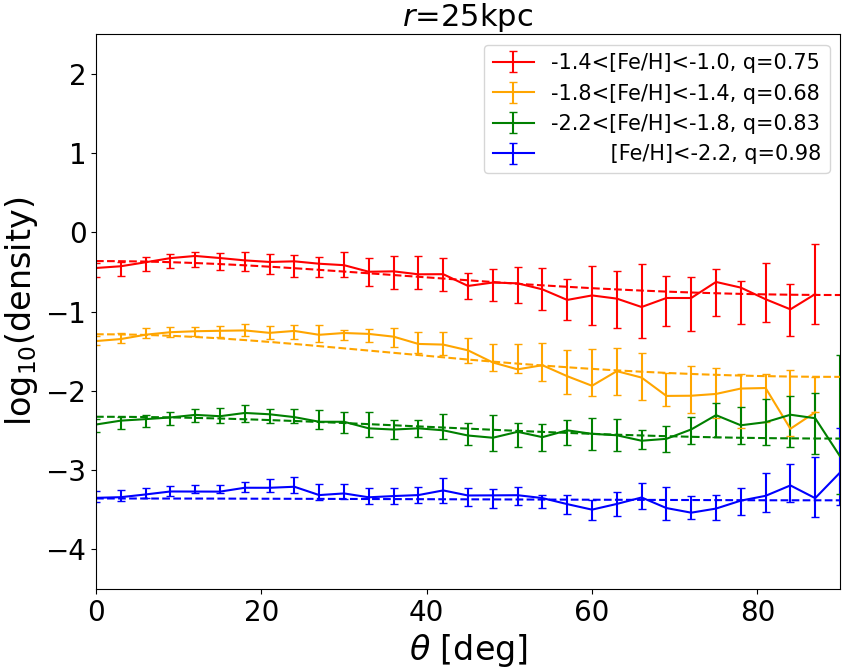}
\caption{The density distribution along $\theta=0^\circ-90^\circ$ at $r=8, 15, 25$~kpc (solid lines) and the fitted curves with the ellipsoidal profile, Eq. \eqref{eq:sing_ellip} (dashed lines) for each metallicity range.
Each curve at a different $r$ is shifted vertically for the comparison between different metallicity ranges.
\label{fig:gene_fit}}
\end{figure}

Figure \ref{fig:gene_pla} shows the reproduced density distribution along the Galactic plane (solid lines) and fitting curves (dashed lines).
We find that all the distributions have a discontinuity at $R\sim8$~kpc,
and the density below this radius falls short of the fitting curves in all stellar metallicity ranges.
The cause of this continuity is due to an observational bias \citep{SLZ1990}:
a star moving only within the region inside the position of the Sun, $r\sim8$~kpc,
i.e., a star with apocentric radius $r_{\rm apo} \lesssim 8$~kpc,
cannot reach the observable region near the Sun.
Therefore, the observational data are devoid of these stars.
This bias is also seen in Figure \ref{fig:ELz_sau}, where stars with $E<-1.5\times 10^5~{\rm km^2~s^{-2}}$,
i.e., those orbiting inside the solar position, are lacking in the current local sample \citep[see also,][]{myeong2018b,carollo2021}.
We thus avoid the innermost distribution $r<8$~kpc for the following fittings.

We first determine the value of the slope $\alpha$
by the root-mean-squares fitting of the radial distributions in Figure \ref{fig:gene_pla} into the power-law function, $\rho\propto R^{-\alpha}$,
i.e. the density distribution at $z=0$ in Eq. \eqref{eq:sing_ellip}.
We obtain $\alpha =3.2-3.4$ in the entire metallicity ranges.
Specifically, the value of $\alpha$ in each metallicity range is $3.42^{+0.17}_{-0.07}$, $3.23^{+0.11}_{-0.08}$, $3.34^{+0.11}_{-0.11}$, and 3$.42^{+0.09}_{-0.13}$, from metal-rich to metal-poor ranges. 

\begin{figure}[t]
\centering
\includegraphics[width=80mm]{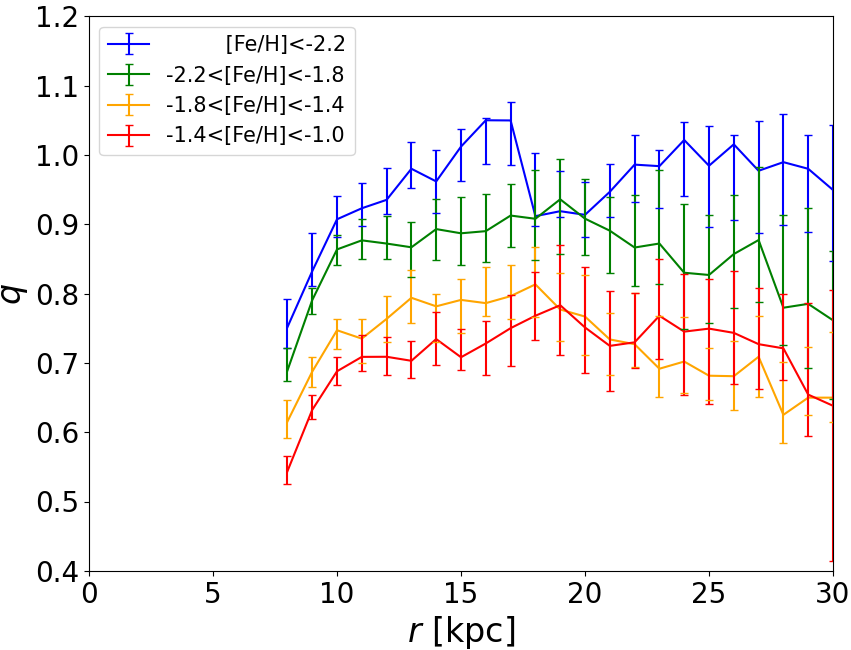}
\caption{The axial ratio $q$ of the ellipsoidal profile \eqref{eq:sing_ellip} as a function of the Galactocentric distance $r$ for each metallicity.
\label{fig:q_gene}}
\end{figure}

Next, we determine the value of an axial ratio, $q$, at each position $r$ and metallicity range.
Figure \ref{fig:gene_fit} shows the reproduced density distribution along $\theta =0^\circ - 90^\circ$ at $r=8, 15, 25$~kpc (solid lines).
The dashed lines correspond to the ellipsoidal fitting curves to these density distributions with a parameter, $q$: 
while these lines be constant along $\theta$ in the case of
a spherically symmetric case of $q=1$, 
a smaller $q$, i.e., more flattened axial ratio, leads to a more negative density gradient with increasing $\theta$.
We find that the stellar halo in a more metal-poor range has a rounder shape,
and metal-rich range is more flattened.
This result is in agreement with previous works before {\it Gaia} using much smaller numbers of sample stars \citep{chiba2000, carollo2007}.

Figure \ref{fig:q_gene} shows the radial distribution of $q$ derived in this manner over $r=8$ to 30~kpc.
We find that more metal-rich halo stars tend to have a smaller axial ratio $q$ over entire radii 
(except for somewhat reversed trend in the ranges of $-1.8<{\rm [Fe/H]}<-1.0$ at $r>20$~kpc),
and that the axial ratio of the most metal-poor halo (${\rm [Fe/H]}<-2.2$) is nearly 1 over entire radii, i.e., having a nearly spherical shape.

\begin{figure}[t]
\centering
\includegraphics[width=80mm]{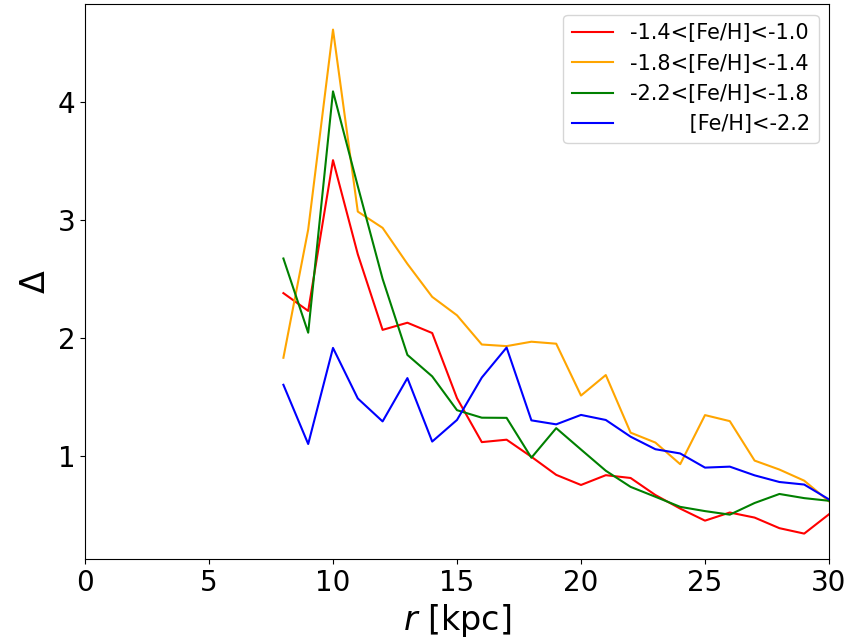}
\caption{The degree of derivation, $\Delta$, defined in Eq. \eqref{eq:Del}, between the reproduced density distribution \eqref{eq:sumrho} and the fitting ellipsoidal profile \eqref{eq:sing_ellip}, normalized by each uncertainty.
\label{fig:chi_gene}}
\end{figure}

It is worth remarking from Figure \ref{fig:gene_coar} and \ref{fig:gene_fit} that the actual shapes of equidensity contours deviate from ellipsoidal profiles in Eq. (\ref{eq:sing_ellip}):
they appear somewhat boxy or peanut-like shapes, so that the derived density distributions depicted in Figure \ref{fig:gene_fit} have a peak around $\theta\sim15^\circ$, whereas simple ellipsoidal profiles should decrease monotonically with $\theta$. 
We note that these properties of the density shapes are related to the lack of halo stars with small $I_3$ or small orbital angle $\theta_{\rm max}$ shown in Figure \ref{fig:EI3} and \ref{fig:histth0} (see Subsection \ref{subsec:412}).
We quantify the degree of the derivation from ellipsoidal shapes by $\Delta$ defined as

\begin{equation}
\Delta(r)=\sqrt{\frac{1}{N_\theta}\sum_{\theta_i}\frac{[\log_{10}\rho_{\rm obs}(r, \theta_i)-\log_{10}\rho_{\rm fit}(r, \theta_i)]^2}{\sigma^2_{\log_{10}[\rho_{\rm obs}(r, \theta_i)]}}}
\label{eq:Del}
\end{equation}
where $\rho_{\rm obs}$ is the density calculated from Eq. \eqref{eq:sumrho} (solid lines in Figure \ref{fig:gene_fit}), $\rho_{\rm fit}$ is the fitted ellipsoid given in Eq. \eqref{eq:sing_ellip} (dashed lines),
and $N_\theta=31$ is the number of bins for $\theta_i$.
Figure \ref{fig:chi_gene} shows the distributions of $\Delta$ having a peak sharply at $r=10$~kpc except for ${\rm [Fe/H]}<-2.2$;
the latter range shows spherical shapes with little deviation from the ellipsoidal distribution over wide range of $r$.
At $r>10$~kpc, $\Delta$ is decreasing with $r$, where $\Delta$ is largest in $-1.8<{\rm [Fe/H]}<-1.4$.

\begin{figure}[t]
\centering
\includegraphics[width=80mm]{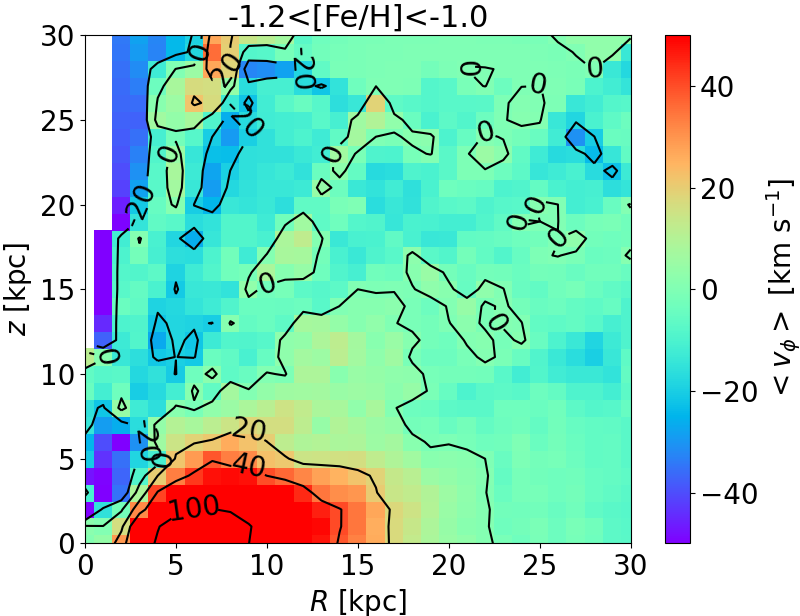}
\includegraphics[width=80mm]{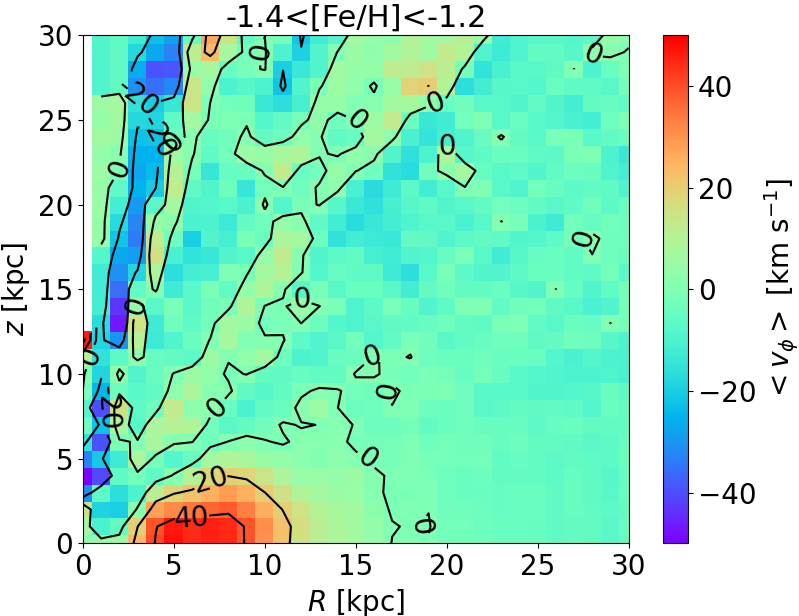}
\includegraphics[width=80mm]{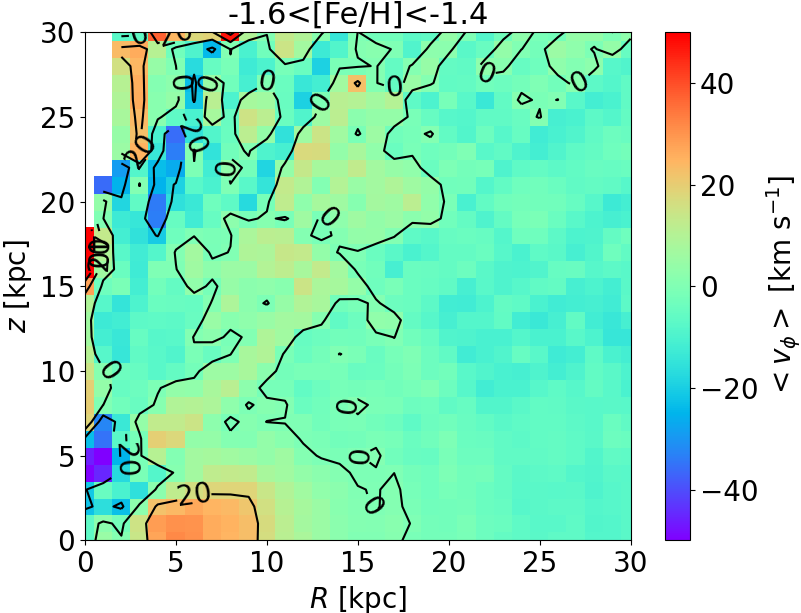}
\caption{
The global distribution of $\left<v_\phi\right>$, defined in Eq. \eqref{eq:muvtau} and \eqref{eq:vphix}, for the halo sample with $-1.2<{\rm [Fe/H]}<-1.0$ (top), $-1.4<{\rm [Fe/H]}<-1.2$ (middle), and $-1.6<{\rm [Fe/H]}<-1.4$ (bottom panel).
As is clear, there exists a region with disk-like kinematics (as realized from red color) corresponding to the metal-weal thick disk.
\label{fig:gene_vphi}}
\end{figure}

These behaviors in Figure \ref{fig:chi_gene} suggest that some substructures that perturb the ellipsoidal profile exist at around $r=10$ kpc in ${\rm [Fe/H]}>-2.2$
and $r>10$~kpc in $-1.8<{\rm [Fe/H]}<-1.4$.
This may include the presence of the metal-weak thick disk (MWTD) \citep[e.g.][]{beers2002, carollo2019} at $r<$10~kpc and/or the presence of Monoceros Stream \citep{newberg2002, yanny2003, ibata2003}.
Alternatively, a large $\Delta$ may suggest that the shape of the stellar halo is intrinsically boxy or peanut-like shaped rather than spheroid.
We discuss the origin of this characteristic shape of the halo in \ref{subsec:412} in more detail.

\subsubsection{Velocity Distribution}
Next, using Eq. \eqref{eq:muvtau},
we obtain the global rotational velocity distribution, $\left<v_\phi\right>$.
The results for relatively metal-rich ranges are shown in Figure \ref{fig:gene_vphi}.
We find that stars in $-1.2<{\rm [Fe/H]}<-1.0$ have the relatively large rotation, whose maximum is over $100~{\rm km~s^{-1}}$.
This disk-like structure is local, $R\lesssim10$~kpc and $z\lesssim3$~kpc, and decays with lowering stellar metallicity.
It is consistent with the properties of the MWTD.
Outside the disk-like region, the velocity field is almost 0, with no significant prograde and retrograde rotation.

\begin{figure}[t]
\centering
\includegraphics[width=80mm]{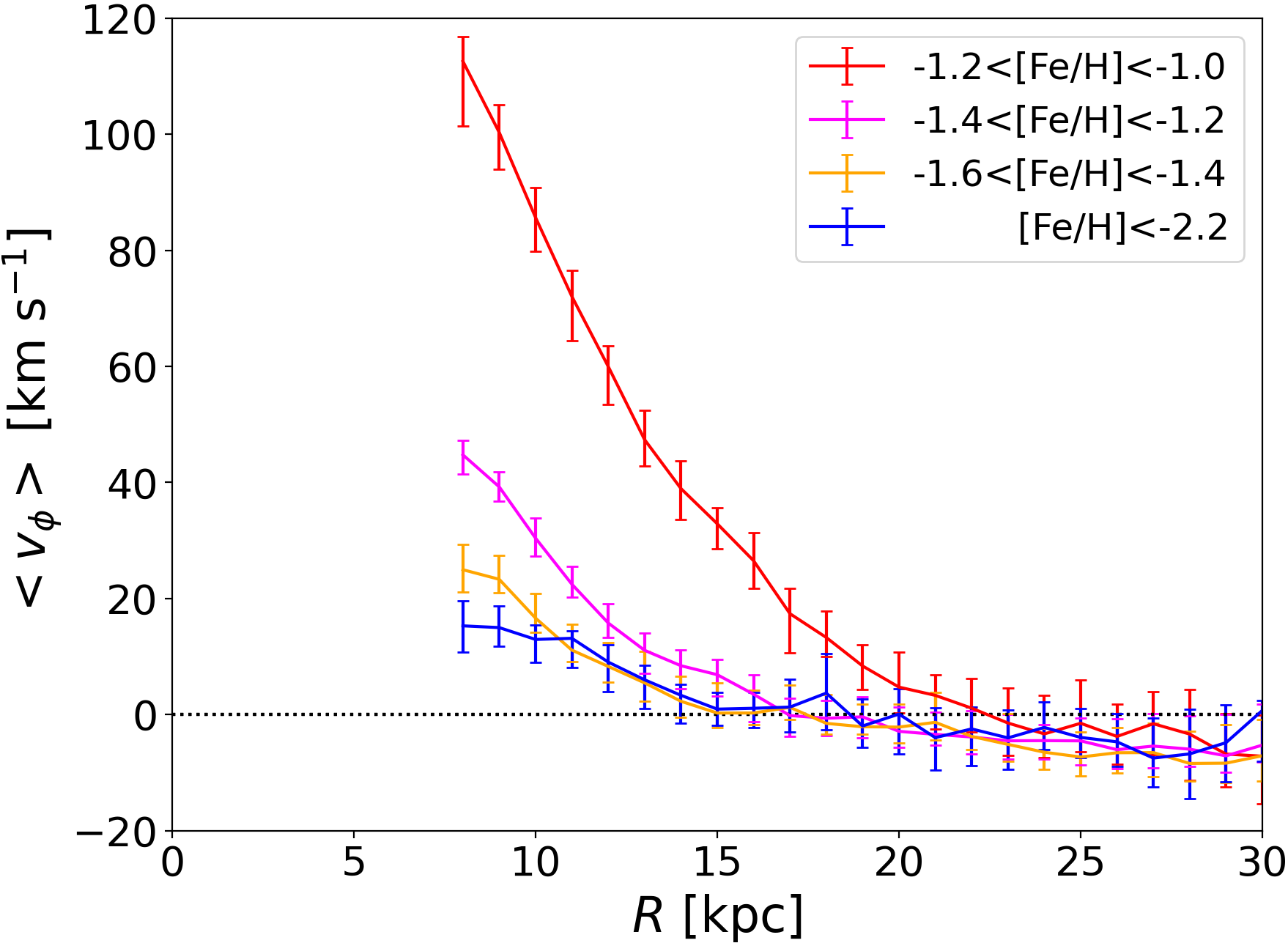}
\caption{The mean rotational velocity along the Galactic plane for each metallicity range.
The dotted line indicates the zero velocity.
\label{fig:vphi_plane}}
\end{figure}

Figure \ref{fig:vphi_plane} shows the rotational velocity distribution along the Galactic plane $z=0$.
The mean rotational velocity in $-1.2<{\rm [Fe/H]}<-1.0$ is more than twice as ones of other metallicity ranges at $8<R<15$~kpc. On the other hand,
over $R>20$~kpc, $\left<v_\phi\right>$ is almost zero in any metallicity ranges.

\begin{figure*}[t]
\centering
\includegraphics[width=80mm]{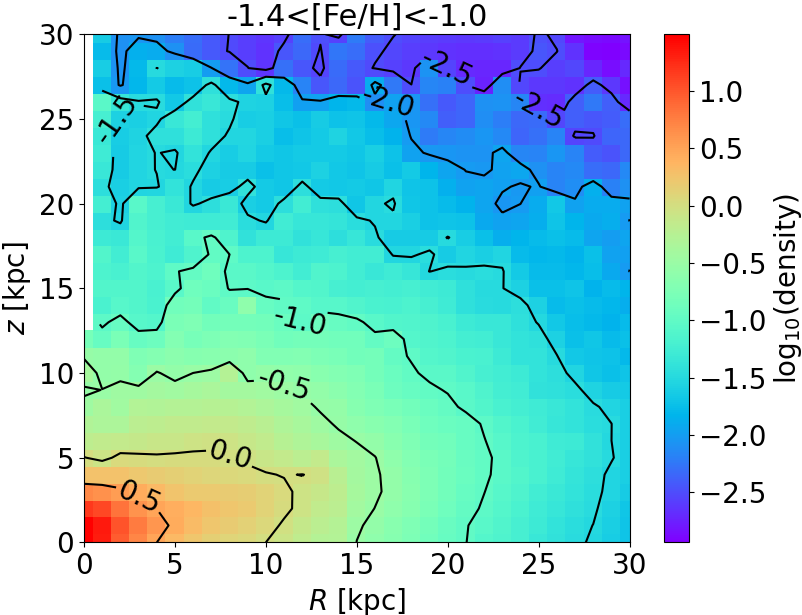} \includegraphics[width=80mm]{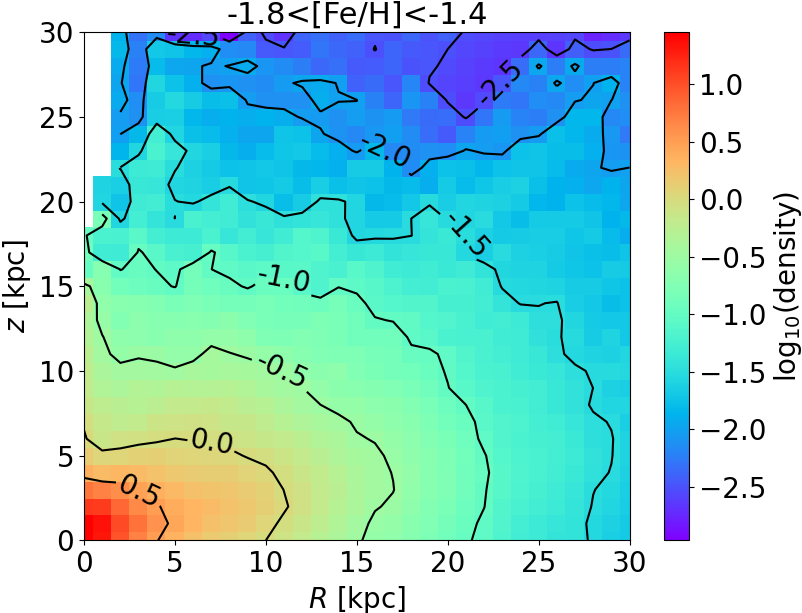}
\includegraphics[width=80mm]{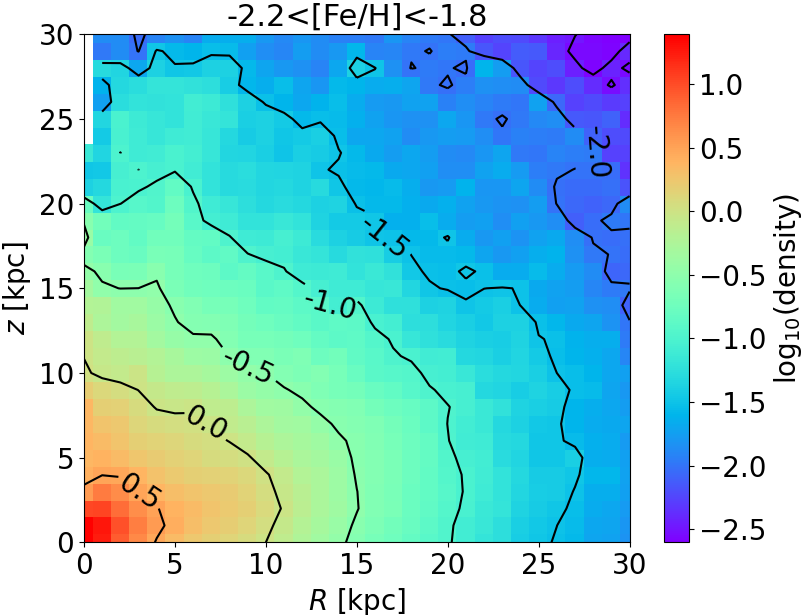} \includegraphics[width=80mm]{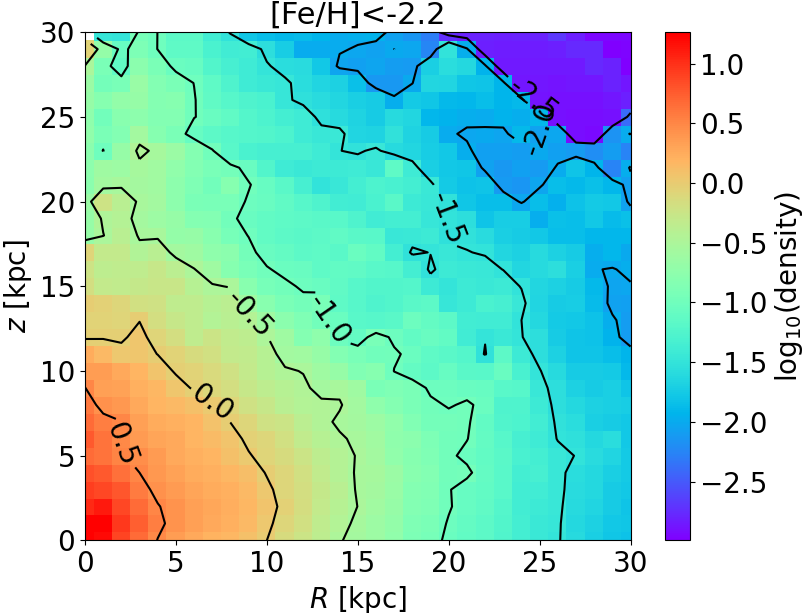}
\caption{The contour of the global density distribution of the GSE-like subsample.
The normalization and the metallicity ranges are
same as in Figure \ref{fig:gene_coar}.
\label{fig:gse_coarse}}
\end{figure*}

Note that Fig. \ref{fig:vphi_plane} shows the mixture distribution of the stellar disk and the stellar halo.
\citet{carollo2019} showed that the rotational velocity of the MWTD reaches $\sim150~{\rm km\ s^{-1}}$,
while Fig. \ref{fig:vphi_plane} shows that the maximum velocity is $\sim110~{\rm km\ s^{-1}}$.
This difference is due to the contamination of halo stars, which move relatively randomly with less rotation.
Thus Fig. \ref{fig:vphi_plane} underestimates the velocity distribution of the pure MWTD.
For the same reason, we do not claim that the metallicity of the MWTD is only over [Fe/H]~$>$~-1.2.

Additionally, Fig. \ref{fig:gene_vphi} and \ref{fig:vphi_plane} show that for the metal-poor stars, especially in ${\rm [Fe/H]}<-2.2$, the velocity field in the outer region, $R>15-20$~kpc, is slightly but systematically negative below zero about $1\sigma$ uncertainty, 
but this feature is not thought to be real for the MW's stellar disk. 
This rotational property for very metal-poor stars in the outer region is in agreement with the retrograde rotation of the outer-halo sample stars \citep[e.g.][]{carollo2010}.
While the sample of previous researches shows a significant retrograde rotation at the solar position, 
the rotational value at $R\sim30$~kpc in this work is reduced at the currently derived one under the angular-momentum conservation.

\subsection{The GSE-like subsample} \label{subsec:GSE}
We carry out the same analysis for the subsample stars showing GSE-like kinematics.
The derived density distributions for this subsample are shown in Figure \ref{fig:gse_coarse}, for the different metallicity ranges, which are the same as in Figure \ref{fig:gene_coar}.
We fit this distribution to the ellipsoidal profile given in Eq. \eqref{eq:sing_ellip}.

\begin{figure}[t]
\centering
\includegraphics[width=80mm]{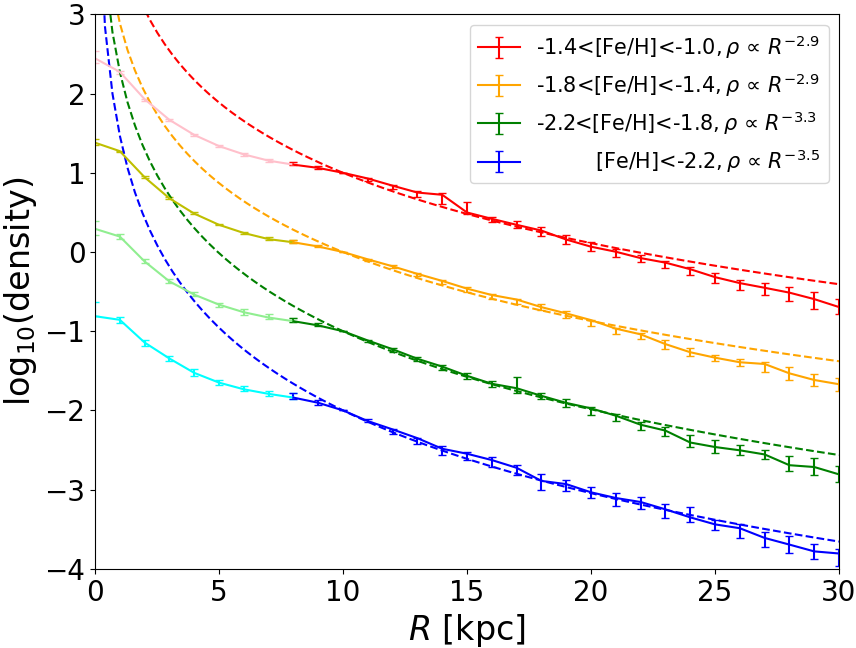}
\caption{The distribution of the GSE-like subsample along the Galactic plane (solid lines) and their fitting curves with a power law (dashed lines),
to be compared with Figure \ref{fig:gene_pla} for the entire halo sample.
This comparison indicates that the density distribution of the GSE-like subsample shows a noticeable drop at $R\sim20$~kpc in the metallicity range of ${\rm [Fe/H]}>-1.8$.
\label{fig:gse_plane}}
\end{figure}

Figure \ref{fig:gse_plane} shows that the density distribution along the Galactic plane (solid lines),
from which we can determine the slope $\alpha$ given in Eq. \eqref{eq:sing_ellip}
for the ellipsoidal profile (dashed lines).
We note that in Figure \ref{fig:gse_plane} there are discontinuities at $R\sim8$~kpc, 
i.e., at the same position in Figure \ref{fig:gene_pla}.
Since these discontinuities can be due to the observational bias, as is the case in the entire halo sample,
we do not use these inner radii to determine the value of $\alpha$.

The determined values of $\alpha$ are $2.94^{+0.16}_{-0.05}$, $2.89^{+0.13}_{-0.00}$, $3.28^{+0.15}_{-0.03}$, and $3.47^{+0.24}_{-0.08}$, respectively, from metal-rich to metal-poor range.
Compared with the case of the entire halo sample, the slopes in relatively metal-rich range, ${\rm [Fe/H]}>-1.8$, are systematically smaller,
and those in more metal-poor range, ${\rm [Fe/H]}<-1.8$, are almost the same.
This result is consistent with the fact that GSE appears only in a relatively metal-rich range \citep{belokurov2018}. 
The small $\alpha$ means that the density distribution of the GSE-like subsample decreases with Galactocentric distance more slowly the general stellar halo.

It is worth noting in Figure \ref{fig:gse_plane} that 
the derived density distribution falls short of a single power-law distribution at around $R=20$~kpc in the relatively metal-rich range, ${\rm [Fe/H]}>-1.8$.
This feature does not appear in the entire halo sample shown in Figure \ref{fig:gene_pla}.
We note that we do not artificially remove the region at $R>20$~kpc when we fit the global distribution and determine $\alpha$.
Thus we conclude that the density distribution of the GSE-like sample shows a drop at around 20~kpc,
which may correspond to the GSE's radial boundary (see Subsection \ref{subsec:gse_distri}).

\begin{figure}[t]
\centering
\includegraphics[width=80mm]{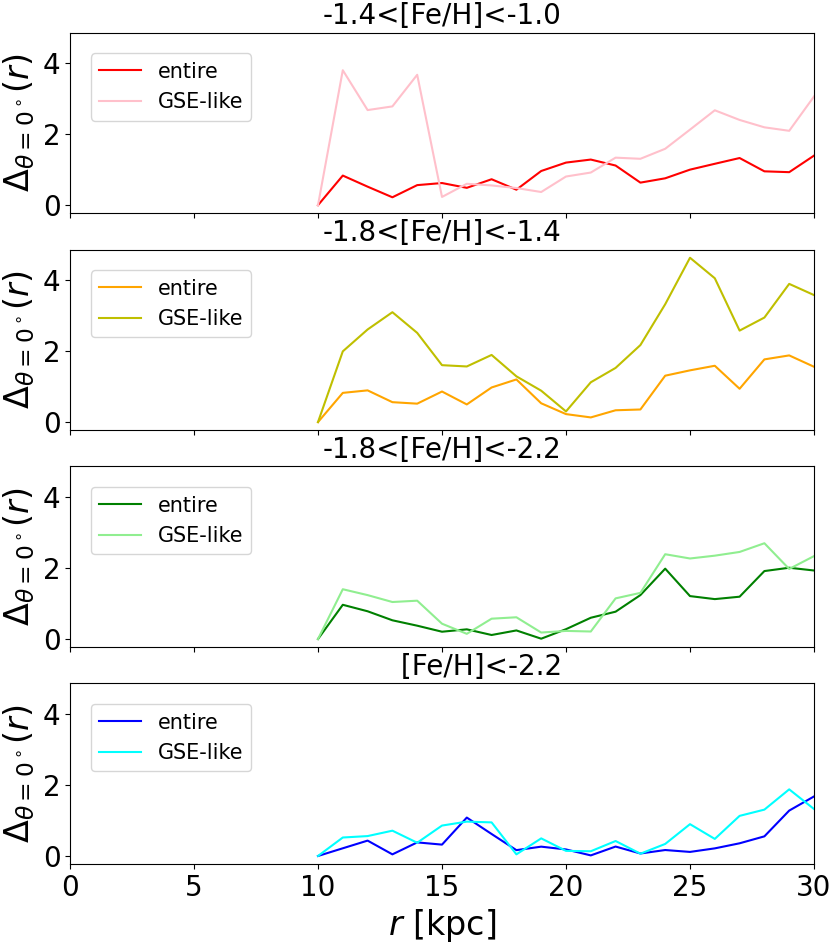}
\caption{The degree of the deviation of the reconstructed density distribution from the ellipsoidal fitting profile along the Galactic plane, $\Delta_{\theta=0^\circ}(r)$ defined by Eq. \eqref{eq:Del_th}.
\label{fig:Del_plane}}
\end{figure}

To clarify that the drops are real properties of GSE-like subsample, we calculate the value indicating the deviation of the reconstructed density distribution from the ellipsoidal profile, defined as
\begin{equation}
\Delta^2_\theta(r)=\frac{[\log_{10}\rho_{\rm obs}(r, \theta)-\log_{10}\rho_{\rm fit}(r, \theta)]^2}{\sigma^2_{\log_{10}[\rho_{\rm obs}(r, \theta)]}}
\label{eq:Del_th}
\end{equation}
This is the same as the factors of the sum in Eq. \eqref{eq:Del}.
We calculate both $\Delta_{\theta=0^\circ}(r)$ for entire halo sample and GSE-like subsample and compare them in Figure \ref{fig:Del_plane}.
We find that $\Delta_{\theta=0^\circ}(r)$ of GSE-like subsample in -1.8$<$[Fe/H]$<$-1.0 gets a remarkably large value at $10\lesssim r\lesssim15$~kpc and $r\gtrsim20$~kpc.
This figure also suggests that the GSE-like subsample in -1.8$<$[Fe/H]$<$-1.0 range does not follow the power-law profile at $r\gtrsim20$~kpc, so this is GSE's characteristic length.
On the other hand, the reason of large $\Delta_\theta(r)$ at $10\lesssim r\lesssim15$~kpc is due to the non-ellipsoidal shape like boxy or peanut-like shape, shown in Figure \ref{fig:chi_gene}.

Next, we determine the value of the axial ratio $q$ for this subsample.
The result is shown in Figure \ref{fig:q_gse}, compared with the case of the entire halo sample.
We find that in the relatively metal-rich range ${\rm [Fe/H]}>-1.8$, 
the axial ratios of the GSE-like subsample are systematically larger than those of the general stellar halo at $r<15$~kpc,
whereas $q$ of the GSE-like subsample are smaller beyond $r\sim20$~kpc.
This is consistent with the Figure \ref{fig:gse_plane}, implying the presence of the radial boundary of the GSE-like members as 20~kpc.
On the other hand, the axial ratios for more metal-poor ranges are larger than or almost the same as compared with ones of the general stellar halo.
Their larger axial ratios may come from the dynamically selection $L_z\sim0$.

Finally, we remark on the $\left<v_\phi\right>$ distribution of this GSE-like sample.
We also find no remarkable structures like the MWTD even in metal-rich ranges,
because of the pre-selection of this sample with $L_z$ around 0.

\section{Discussion} \label{sec:dscs}
Derived global structures of the current halo sample based on the superposition of their orbits suggest several implications for the formation and evolution histories of the MW's stellar halo.
We discuss here what insights are inferred from the current analysis in comparison with recent numerical simulations for the formation of stellar halos.

\subsection{Global halo structure from the entire sample} \label{subsec:whole_halo}
\subsubsection{Implication from axial ratios and radial profiles} \label{subsec:411}
As clearly shown in Figure \ref{fig:q_gene},
the global shapes of relatively metal-rich stellar halos in the range of ${\rm [Fe/H]}>-1.8$ have smaller axial ratios $q$, i.e. more flattened, than more metal-poor halos with ${\rm [Fe/H]}<-1.8$ over all radii.
Also, at radii below 18~kpc, we find that a more metal-rich halo shows a more flattened shape for all the metallicity ranges below ${\rm [Fe/H]}=-1$.

\begin{figure}[t]
\centering
\includegraphics[width=80mm]{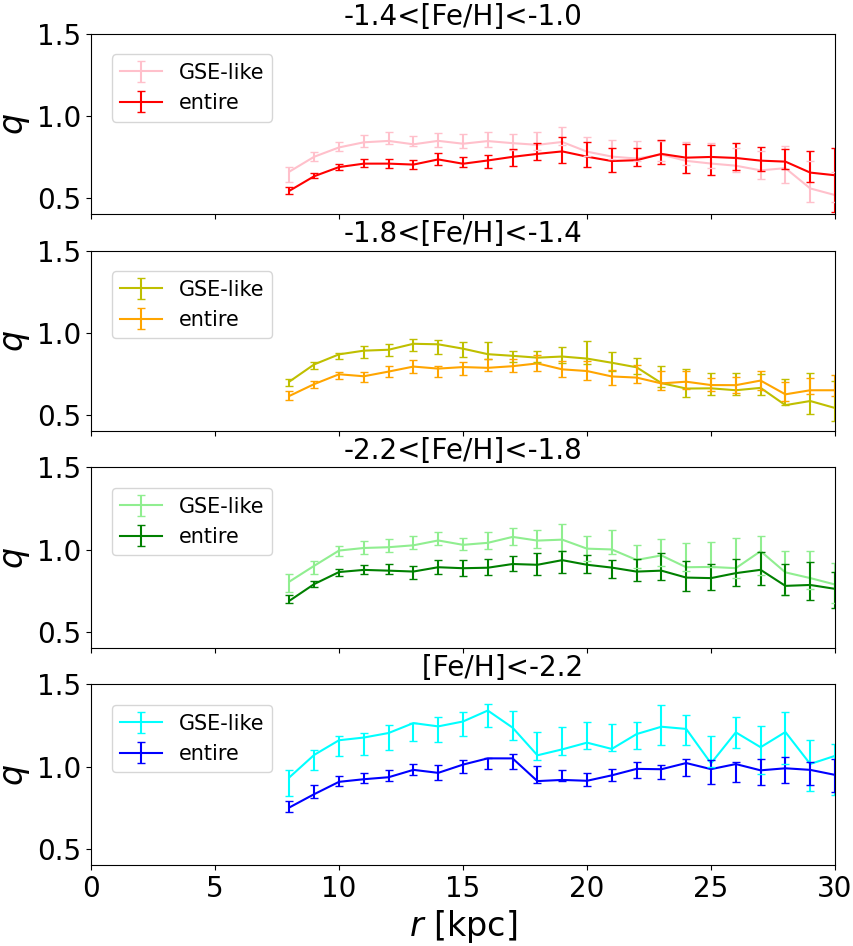}
\caption{The axial ratios $q$ of the GSE-like subsample compared with those of the entire halo sample as a function of the Galactocentric distance $r$ for each metallicity range.
\label{fig:q_gse}}
\end{figure}

The flattened structures in relatively metal-rich ranges may, at least partly, be affected by the presence of the MWTD at ${\rm [Fe/H]}<-1$.
The disk-like motion of the MWTD is appreciable in the $\left<v_\phi\right>$ distribution of $-1.2<{\rm [Fe/H]}<-1.0$ range (Figure \ref{fig:gene_vphi} and \ref{fig:vphi_plane}), with the radial and vertical sizes of $\sim14$~kpc and vertical of $\sim4$~kpc, respectively.
If we adopt a condition, $\left<v_\phi\right>/\sigma_\phi>1$, with $\sigma_\phi=50~{\rm km~s^{-1}}$ as a typical velocity dispersion of the thick disk in $\phi$ direction \citep{chiba2000},
these figures also show that such a disk-like motion becomes weak quickly in $-1.4<{\rm [Fe/H]}<-1.2$, 
where $\left<v_\phi\right>/\sigma_\phi$ is smaller than 1 at all radii,
so the effect of the MWTD may be sufficiently weak at ${\rm [Fe/H]}$ between $-1.4$ and $-1.2$.
The fact that the derived relatively metal-rich halo sample shows a flattened shape over the regions beyond this MWTD-like feature permeates that this flattened structure is a real halo structure.
These flattened parts of the halo may correspond to the {\it in-situ} halo, in which stars have been formed inside a main parent halo.

The formation of this flattened halo parts may be explained within the framework of the CDM model \citep{white1978, bekki2000}:
a flattened structure of a relatively metal-rich halo was formed by dissipative merging of few massive clumps at early stage of the MW's evolution.
Alternatively, this flattened structure can be a dynamically heated, pre-existing disk driven by falling satellites
\citep[e.g.][]{benson2004, purcell2010, maccarthy2012, tissera2013, cooper2015}.

On the contrary, more metal-poor halos hold a structure closer to a sphere,
and this characteristics suggests that it is the {\it ex-situ} halo component.
The presence of this metal-poor, spherical halo is consistent with the MW's formation scenario under the CDM model:
the outer halo was formed by the later accretion of small clumps resembling metal-poor dSphs,
which are stripped by tidal force from the MW's dark halo \citep{bekki2001, carollo2007}.

Recent numerical simulations based on CDM models provide similar chemodynamical properties of stellar halos to those reported here.
\citet{monachesti2019} simulated the formation process of MW-like stellar halos with accretion of stars.
It showed that the axial ratios of the halos are typically 0.6 to 0.9 at $R=20-40$~kpc.
This simulated range is consistent with our result shown in Figure \ref{fig:q_gene}.
It also showed that the shapes of MW-like stellar halos can be prolate in outer regions, where the {\it ex-situ} component dominates.
Our result of ${\rm [Fe/H]}<-2.2$ range has the same shape,
so it is considered that the {\it ex-situ} component is dominant in this metallicity range.

Regarding the radial density slope, $\alpha$, the derived values in this work are almost the same, $3.2-3.4$, in different metallicity ranges
though there may exist varieties in the formation processes.
\citet{rodriguez2016} and \citet{monachesti2019} showed that the accreted halo components have a shallower density slope with $\alpha<3.0$, in the radius range $15<r<100$~kpc than the {\it in-situ} components.
On the other hand, in the range $8<r<30$~kpc
that the current work is concerned, the {\it ex-situ} halo should be less dominant than
in the range of $15<r<100$~kpc.
Thus the values of $\alpha$ may not be strongly affected by later accretion process of small clumps
and this work obtain the almost same values of $\alpha$ for different metallicity ranges.

\subsubsection{Implication from a boxy/peanut-shape configuration} \label{subsec:412}
Here we discuss the origin of the non-ellipsoidal profile peaked at around $r\sim10$~kpc, as shown in Figure \ref{fig:gene_coar}.
We note that the presence of the MWTD is insufficient to explain this feature,
because the non-ellipsoidal structure is most prominent within $-1.8<{\rm [Fe/H]}<-1.4$,
whereas the MWTD dominates in $-1.4<{\rm [Fe/H]}<-1.0$ (Figure \ref{fig:gene_vphi} and \ref{fig:vphi_plane}).
We consider two possibilities to explain this feature.

First, the intrinsic structure of the inner part of the MW's stellar halo forms a boxy/peanut-like shape rather than the ellipsoidal at $r<20$~kpc.
\citet{carollo2007, carollo2010} showed that the inner halo component is dominant up to $r=10-15$~kpc, and the outer halo is dominant at $r>20$~kpc.
These regions correspond to those, where the current deviation measure from an ellipsoidal shape, $\Delta$, shows a peak and is relatively small, respectively.

Figure \ref{fig:gene_fit} shows that the density distributions along $r=8$ and 15~kpc is curved upward at around $\theta\sim15^\circ$ and downward over $\theta>60^\circ$.
In fact, the density deficiency at large $\theta$ also appeared in the results of \citet{SLZ1990}, \citet{chiba2000}, and \citet{carollo2007}.
It was interpreted due to the observational bias:
there is a small probability that such stars orbiting at large $\theta$ are represented in the solar neighborhood and this effect may have been non-negligible in previous works using only a small number of sample stars available before {\it Gaia}.
However, our reconstruction using much numerous stars cataloged in {\it Gaia} leads the similar shape of the density distribution of the MW's stellar halo.
There are indeed no discontinuities in the observational inclination distribution shown in  Figure \ref{fig:histth0}.
Thus we regard the boxy/peanut-shaped structure as a real feature, not biased.

It is also worth remarking in Figure \ref{fig:histth0} that there is a sharp drop in the number of orbits at $\theta_{\rm max}<10^\circ$.
\citet{carollo2021} showed that the lack of these low-$\theta$ (or low-$I_3$) is not artificial:
if stars with small orbital inclination really exist in the MW,
such stars surely pass by the solar neighborhood and thus can be included in the current local sample.
This fact means that since such low-$\theta$ stars possess highly eccentric, radial orbits,
they can surely be identified, if they exist, in the current survey volume.
Then, the lack of such stars near the Galactic plane leads to the reconstructed global density distribution having a boxy shape.

Second, the non-ellipsoidal structure of the halo may reflect the Monoceros Stream \citep{newberg2002, yanny2003, ibata2003}.
This stream feature is a ring-shaped overdensity with heliocentric radius of 6~kpc in the south and 9~kpc in the north \citep{morganson2016} 
and height of $3<z<4$~kpc \citep{ivezic2008}.
These spatial distributions generally agree with our results that the density contours in Figure \ref{fig:gene_coar} are curved upward to the similar vertical range and the non-ellipsoidal structure as shown in Figure \ref{fig:chi_gene} is dominant over the similar Galactocentric distance.

However, \citet{ivezic2008} and \citet{morganson2016} showed that Monoceros Stream is centered at ${\rm [Fe/H]}=-0.95$ and spreads $-1.1<{\rm [Fe/H]}<-0.7$.
This metallicity is quite different from the range where the non-ellipsoidal structure dominates in Figure \ref{fig:chi_gene}, $-1.8<{\rm [Fe/H]}<-1.4$.
This suggests that the non-ellipsoidal structure of the stellar halo obtained here is not due to the Monoceros Stream.

We thus conclude that the non-ellipsoidal structure of the stellar halo, which mainly reveals in the inner parts of the halo, is a physically real feature and
this shape of the stellar halo is considered to reflect the MW's merger history.
It is known that a merger between early-type gas-poor galaxies with same masses (dry and major merger) forms a boxy shaped elliptical galaxy \citep{naab2006, bell2006, cox2006}.
Even including gas in merging process,
while gas dissipates into the  equatorial plane and form more metal-rich stars with disk-like kinematics,
pre-existing stars in progenitor satellites can end up with a boxy structure after merging,
which consists of 3D box orbits.
Also, dynamical heating of this formed disk through later merging of satellites is another source of halo stars,
where heating process can be more effective in the outer disk regions due to weaker vertical gravitational force,
so that the heated debris would look like a peanut-shaped structure,
like edge-on view of long-lived bars \citep{athanassoula2005}.
Indeed, the simulation works under CDM scenarios show some boxy/peanut shaped stellar halos \citep[e.g.][]{bullock2005, cooper2010, monachesti2019}.
Further theoretical works will be needed to quantify this non-ellipsoidal structure of the simulated halos and clarify what merging histories of progenitor galaxies are associated with this characteristic structure.

\subsubsection{On the discontinuities by the observational bias} \label{subsec:inout}
\begin{figure}[t]
\centering
\includegraphics[width=80mm]{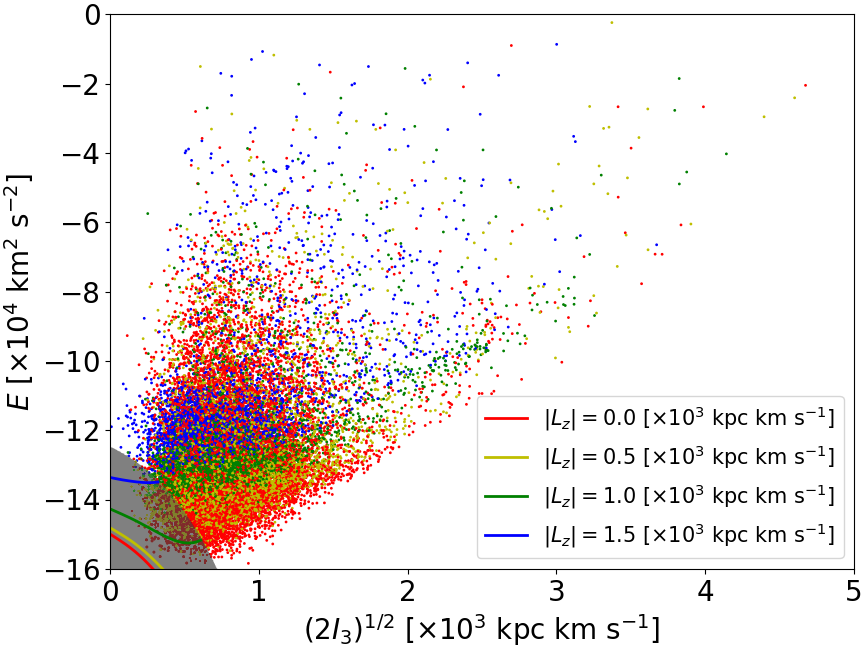}
\caption{The distribution of sample stars in the phase space $(\sqrt{2I_3}, E)$.
The color of the points indicates the vertical angular momentum $|L_z|$ of stars;
red: $0.0<|L_z|<0.5$, yellow: $0.5<|L_z|<1.0$, green: $1.0<|L_z|<1.5$, blue: $|L_z|>1.5$ (
in units of $L_z$ is $\times10^3{\rm kpc\ km\ s^{-1}}$).
The shaded region is the area, where the stars stay only in "the inner region" relative to the survey volume, 
depending on the values of $|L_z|$ depicted with the color-coded lines: 
such stars having $|L_z|$ larger than a specific value locate below the corresponding colored-line within the shaded region.}
\label{fig:inner}
\end{figure}

We have mentioned that the reproduced density distribution has remarkable discontinuities at $R\sim8$~kpc (cf. Fig. \ref{fig:gene_pla} and \ref{fig:gse_plane}).
These discontinuities are caused by the observational bias that stars orbiting only in the inner region of $R\lesssim8$~kpc never reach the observable region near the Sun, and are not contained in the currently adopted stellar catalog.
However, in principle, it is also conceivable that another discontinuity may appear caused by stars staying far from the Sun,
but we do not find such discontinuities.
This subsection explains the properties of the observational bias caused by stars staying inside and outside the locally observable region.

In the following, ``the inner region" is defined as $|z|<(R-R_\odot)\tan165^\circ$ and  ``the outer region" is $|z|<(R-R_\odot)\tan15^\circ$, which respectively are the X-shape vacant regions in Fig. \ref{fig:Rz}.
Stars staying either of these regions are never included the observational sample.

Fig. \ref{fig:inner} shows the $E$-$I_3$ space distribution, similar to Fig. \ref{fig:EI3},
but the color of the points indicates the range of $|L_z|$.
The shaded region indicates the area where stars staying in the inner region locate, depending on their $|L_z|$.
For example, stars staying in the inner region with $|L_z|>1.0\times10^3 {\rm\ kpc\ km\ s^{-1}}$ locate in the shaded region below the green line.
In other words, stars shown with green and blue points (observed stars with $|L_z|>1.0\times10^3 {\rm\ kpc\ km\ s^{-1}}$) cannot locate in the shaded region below the green line due to the bias.
It is clear that the shaded region in Fig. \ref{fig:inner} demonstrates the observational bias in the inner region.

\begin{figure}[t]
\centering
\includegraphics[width=80mm]{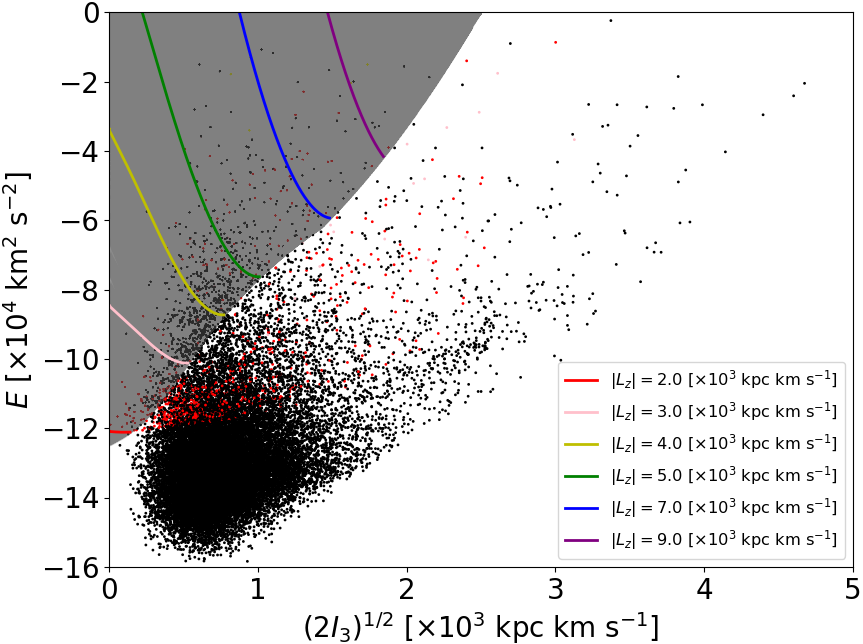}
\caption{The phase-space distribution of sample stars as in Fig. \ref{fig:inner},
but the color of the points corresponds to the different range for $|L_z|$;
black: $0.0<|L_z|<2.0$, red: $2.0<|L_z|<3.0$, pink: $3.0<|L_z|<4.0$, yellow: $|L_z|>4.0$ (in units of $L_z$ is $\times10^3\ {\rm kpc\ km\ s^{-1}}$).
The shaded region is the area, where the stars staying only in "the outer region" relative to the survey volume, 
depending on the values of $|L_z|$ depicted with the color-coded lines: 
such stars having $|L_z|$ larger than a specific value locate below the corresponding-colored line within the shaded region.}
\label{fig:outer}
\end{figure}

Fig. \ref{fig:outer} also shows the $E$-$I_3$ space distribution, but the shaded region indicates the area where stars staying only in the outer region.
In this case, for example, red-pointed stars (with $|L_z|>2.0\times10^3 {\rm\ kpc\ km\ s^{-1}}$) cannot locate in the shaded region below the red line;
this prohibited region is so small.
For another example, stars with $|L_z|>7.0\times10^3 {\rm\ kpc\ km\ s^{-1}}$ are greatly constrained such that they cannot locate in the wide shaded region below the blue line, 
but such stars with high-$|L_z|$ do not exist at least within $R\sim35 $~kpc, provided a circular rotation velocity there remains about $200~{\rm km\ s^{-1}}$ (cf. Fig. \ref{fig:rotcurve} and \ref{fig:ELz_sau})).
Additionally, the black points ($|L_z|<2.0\times10^3 {\rm\ kpc\ km\ s^{-1}}$, 96\% of the sample) are not limited in the $E$-$I_3$ plane,
which means that they cannot stay in the outer region.
It suggests that stars staying only in the outer region are few, so the corresponding discontinuity is not expected to appear.

\citet{carollo2021} supports the current argument.
It showed that the $\sqrt{2I_3}$ distribution quickly damps at around $\sqrt{2I_3}\sim0.5\times10^3 {\rm\ kpc\ km\ s^{-1}}$.
Stars staying in the outer region need to have small inclination $\theta_0$.
Fig. \ref{fig:thI3} shows that $\theta_0$ strongly correlates with $I_3$.
Thus the damp of $\sqrt{2I_3}$ distribution suggests that stars staying in the outer region is very few.
On the other hand, stars staying in the inner region need to have low $E$.
In Fig. \ref{fig:ELz_sau}, the dotted line descend beyond the frame around $|L_z|\sim0-1.0\times10^3 {\rm\ kpc\ km\ s^{-1}}$, but the data points do not follow this lower-limit lines.
The gap between the dotted line and the fiducial points suggests that there are many inner stars not captured by the observation.
Therefore, the discontinuities at around $R\sim8$~kpc appear
while there are no discontinuities in the outer region.

\subsection{Global structure from the GSE-like subsample} \label{subsec:gse_distri}
As shown in Figure \ref{fig:gse_plane},
the distributions of the GSE-like subsample in $-1.8<{\rm [Fe/H]}<-1.0$ fall short from a single power-law profile at $R\sim20$~kpc on the Galactic plane.
This metallicity range is just consistent with the reported metallicities for the likely GSE member stars \citep{belokurov2018, koppelman2019}.
By combining with the result in Fig. \ref{fig:Del_plane},
we regard 20~kpc as GSE's likely radial scale.

\begin{figure}[t]
\centering
\includegraphics[width=80mm]{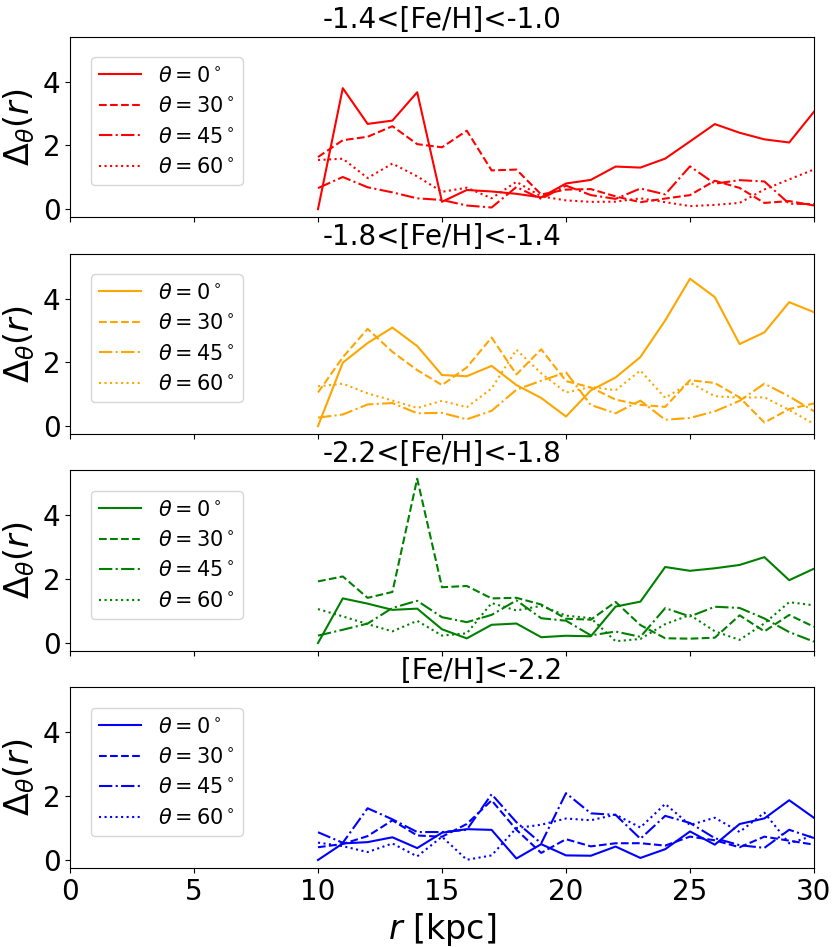}
\caption{The degree of the deviation $\Delta_\theta(r)$ of GSE-like sample along $\theta = 0^\circ, 30^\circ, 45^\circ$ and $60^\circ$,
which correspond to solid, dashed, dashed-dotted, and dotted line, respectively.
\label{fig:Del_th}}
\end{figure}

The result that GSE's scale is 20~kpc is consistent with some previous works. 
\citet{naidu2020} showed the occupancy fraction of each substructure as a function of position based on the observation of H3 survey,
and revealed that GSE's fraction peaks at $r\sim20$~kpc and rapidly declines with increasing $r$.
\citet{deason2018} found the sausage-like structure in the stellar halo has its apocenter at $25-30$~kpc.
\citet{lancaster2019} also indicated that GSE's density rapidly decreases at 20~kpc.

This scale is considered to indicate the position of the apocenter of GSE's progenitor.
Because it is thought that GSE is a trace of a merger event between the MW and a large dwarf galaxy about 10~Gyr ago, 
the radius 20~kpc should be the characteristic scale of the event.
That ``characteristic scale'' is typically expected the apocenter of the merged galaxy \citep{deason2018, lancaster2019}.

Furthermore, the distance of 20~kpc is consistent with the end where the inner halo is dominant \citep{carollo2007, carollo2010},
so our result suggests that GSE is the dominant substructure within the inner halo.
This inference is supported by the behavior in $r=10-15$~kpc of the GSE-like sample.
We have mentioned that the large values of $\Delta(r)$ might be the feature of the inner halo (Section \ref{subsec:412}), 
and Fig. \ref{fig:Del_plane} shows that  $\Delta_{\theta=0^\circ}(r)$ in $r=10-15$~kpc and $-1.8<{\rm [Fe/H]}<-1.0$ of the GSE-like sample is much larger than one of the entire halo sample. 
Namely, the GSE-like sample reflects the basic properties of the inner halo significantly.

We also attempt to find the characteristic inclination $\theta$ of the GSE-like sample.
Fig. \ref{fig:Del_th} shows the deviation measure $\Delta_\theta(r)$ of the GSE-like sample for various $\theta$.
It shows that three lines except for $\theta=0^\circ$ are similar and have no remarkable difference for each metallicity range.
Exceptionally, $\theta=30^\circ$ line of $-2.2<{\rm [Fe/H]}<-1.8$ has a sharp peak at $R\sim14$~kpc, but it is likely to be an outlier.
Thus, the GSE's characteristic inclination is not available here.

Next, we have shown that the slope $\alpha$ of the GSE-like subsample in $-1.8<{\rm [Fe/H]}<-1.0$ is 2.9, being systematically smaller than slopes of the general halo, $\alpha=3.2-3.4$.
On the other hand, in more metal-poor ranges ${\rm [Fe/H]}<-1.8$, the values of $\alpha$ are almost the same for both samples.
As shown in Subsection \ref{subsec:411}, it is known that accreted halo components have smaller $\alpha$ with $\alpha<3.0$ \citep{rodriguez2016, monachesti2019}.
Thus smaller $\alpha$ of GSE-like subsample in our work is consistent with its origin.

Additionally, Figure \ref{fig:q_gse} shows that the axial ratios of the GSE-like subsample in $-1.8<{\rm [Fe/H]}<-1.0$ are systematically larger than the general stellar halo within $r\lesssim20$~kpc,
suggesting that GSE's shape is closer to a sphere than the general halo.
This property of the GSE, which is a merging origin, was also suggested in the simulation work by \citet{monachesti2019}.
It showed that a stellar halo which experienced accretion of a satellite recently forms more spherical shape than a general halo.

\subsection{The completeness of the observational data} \label{subsec:complete}
This work analyzes the global distribution of the MW's stellar halo by using local observational data.
There might be some concern about the effect of the completeness in the result.
We carry out the additional analysis below to get insights into this issue.

\begin{figure}[t]
\centering
\includegraphics[width=80mm]{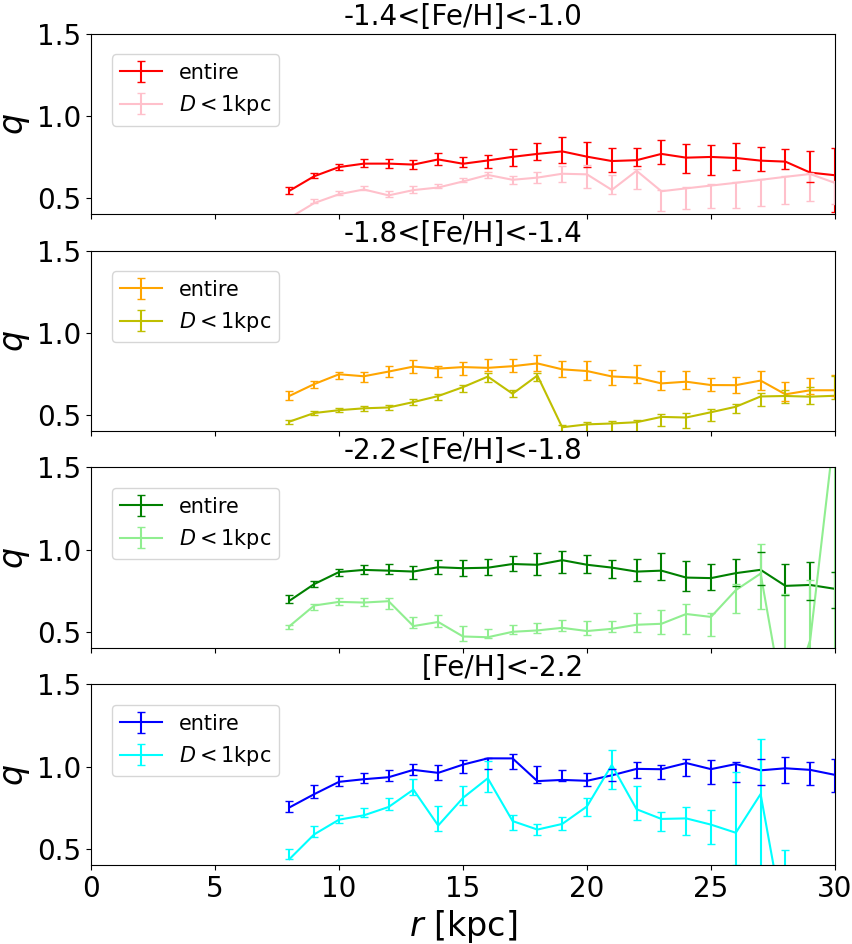}
\includegraphics[width=80mm]{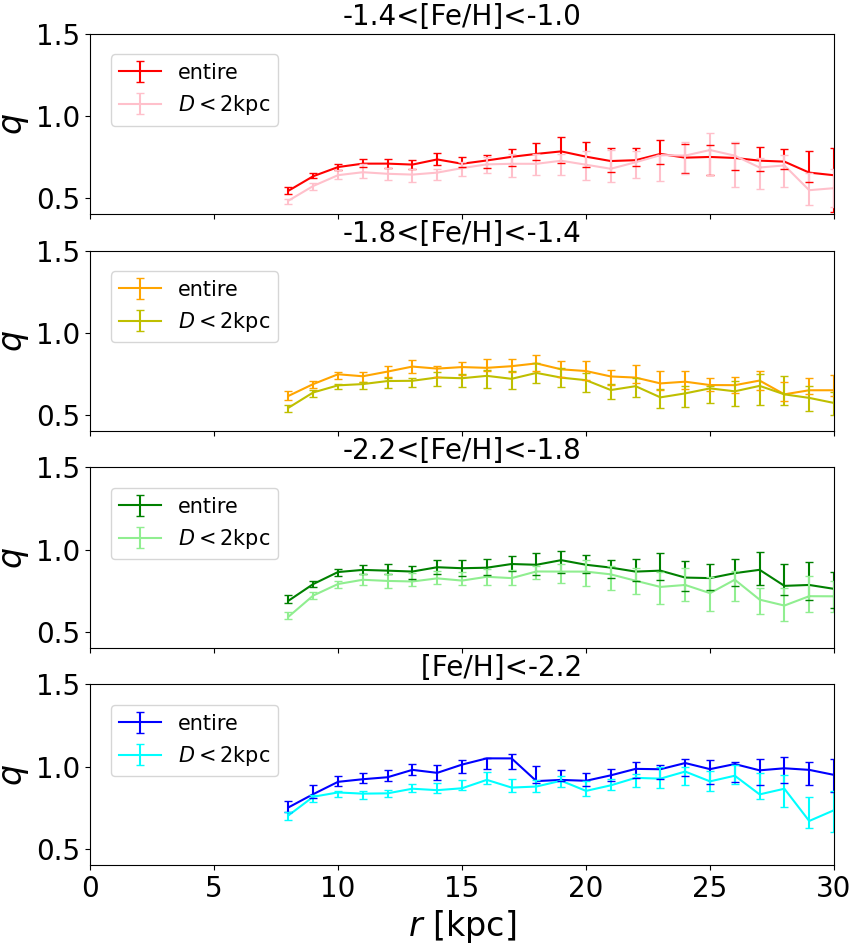}
\caption{The result for the axial ratio of subsamples of $D<1$~kpc (upper panel) and $D<2$~kpc (lower panel), compared with the main sample (dark-colored lines, which are the same as in Fig. \ref{fig:q_gene}.)}
\label{fig:nei12}
\end{figure}

We create other two subsamples, whose additional selection condition is that the heliocentric distance, $D$, is smaller than 1~kpc and 2~kpc, respectively.
We carry out the same analysis for these subsamples.

The axial ratio of these subsample is shown in Fig. \ref{fig:nei12} as the representative result.
In the case of $D<1$~kpc 
the reconstructed distribution is very bumpy with large uncertainty (the upper panel of Fig. \ref{fig:nei12}), so this result has poor reliability.
Two reasons of the poor reliability are considered.
The first reason is that the $D<1$~kpc sample has too few small number of stars (1,895 stars, 7.6\% of the main sample) to reproduce the global distribution.
The second is that the region within 1~kpc from the Sun is too small to represent a global distribution.
In any case, this result has insufficient completeness.

On the other hand, for $D<2$~kpc subsample, 
we obtain the result almost the same as the main result (the lower panel of Fig. \ref{fig:nei12}).
It is suggested that the completeness of $D<2$~kpc subsample is sufficient to reproduce the global distribution up to $\sim30$~kpc.
Therefore, it suggests that our main result, using stellar $D\lesssim4$~kpc, is insensitive to the sample completeness.

\section{Conclusion} \label{sec:cncl}
We have constructed and analyzed the global distribution of the MW's stellar halo using {\it Gaia} EDR3 combined with SDSS DR16 SEGUE catalogs.
For this purpose, we adopt the method developed by \citet{SLZ1990} following \citet{may1986}'s strategy of reconstructing the global halo based on the superposition of stars' orbits.
We calculate the orbital density of each star and assign a proper weight to it through a maximum likelihood approach.
We then sum up all of the orbital densities with these weights and derive the global stellar distribution from observational sample stars currently located within a local volume near the Sun.

We have found that more metal-rich halo stars show more flattened distribution as have been derived in previous studies using much smaller numbers of sample stars \citep[e.g.][]{carollo2007}.
In the metal-rich range of $-1.4<{\rm [Fe/H]}<-1.0$, we have identified, through the reconstruction of global rotational velocities, $\left<v_\phi\right>$, the presence of the MWTD at $z<4$~kpc and $R<15$~kpc.
The detailed configuration of the global halo structures in relatively metal-rich ranges of ${\rm [Fe/H]}>-1.8$ is found to have boxy/peanut-shape rather than a simple ellipsoidal shape and this may reflect a past merging event.
On the other hand, in most metal-poor range of ${\rm [Fe/H]}<-2.2$, the halo structure is close to a spherical shape.

We have selected the subsample of the GSE-like stars with $E>-150,000~{\rm km^2~s^{-2}}$ and $-500<L_z<500~{\rm kpc~km~s^{-1}}$.
In contrast to the entire halo sample, the global density profile of these GSE-like stars show a noticeable drop at $r\sim20$~kpc over $0^\circ<\theta<45^\circ$, implying a apocentric boundary of debris stars from a merging progenitor galaxy of the GSE.
Additionally, the behaviors of $\alpha$ and $q$ of GSE-like subsample indicate the features of accreted halo components:
its density distribution decreases more slowly with Galactocentric radius and has a more spherical shape over $r < 20$~kpc than the general stellar halo sample.

\acknowledgments
We are grateful to the referee for invaluable comments, which help revise the current paper. 
We also thank Daniela Carollo for her many useful comments on the original manuscript. We acknowledge support from the JSPS and MEXT Grant-in-Aid for Scientific Research (No. 17H01101,  18H04434 and 18H05437).

Funding for the Sloan Digital Sky 
Survey IV has been provided by the 
Alfred P. Sloan Foundation, the U.S. 
Department of Energy Office of 
Science, and the Participating 
Institutions. 

SDSS-IV acknowledges support and 
resources from the Center for High 
Performance Computing  at the 
University of Utah. The SDSS 
website is www.sdss.org.

SDSS-IV is managed by the 
Astrophysical Research Consortium 
for the Participating Institutions 
of the SDSS Collaboration including 
the Brazilian Participation Group, 
the Carnegie Institution for Science, 
Carnegie Mellon University, Center for 
Astrophysics | Harvard \& 
Smithsonian, the Chilean Participation 
Group, the French Participation Group, 
Instituto de Astrof\'isica de 
Canarias, The Johns Hopkins 
University, Kavli Institute for the 
Physics and Mathematics of the 
Universe (IPMU) / University of 
Tokyo, the Korean Participation Group, 
Lawrence Berkeley National Laboratory, 
Leibniz Institut f\"ur Astrophysik 
Potsdam (AIP),  Max-Planck-Institut 
f\"ur Astronomie (MPIA Heidelberg), 
Max-Planck-Institut f\"ur 
Astrophysik (MPA Garching), 
Max-Planck-Institut f\"ur 
Extraterrestrische Physik (MPE), 
National Astronomical Observatories of 
China, New Mexico State University, 
New York University, University of 
Notre Dame, Observat\'ario 
Nacional / MCTI, The Ohio State 
University, Pennsylvania State 
University, Shanghai 
Astronomical Observatory, United 
Kingdom Participation Group, 
Universidad Nacional Aut\'onoma 
de M\'exico, University of Arizona, 
University of Colorado Boulder, 
University of Oxford, University of 
Portsmouth, University of Utah, 
University of Virginia, University 
of Washington, University of 
Wisconsin, Vanderbilt University, 
and Yale University.

This work has made use of data from the European Space Agency (ESA)
mission {\it Gaia} (\url{https://www.cosmos.esa.int/gaia}), processed by
the {\it Gaia} Data Processing and Analysis Consortium (DPAC,
\url{https://www.cosmos.esa.int/web/gaia/dpac/consortium}). Funding
for the DPAC has been provided by national institutions, in particular
the institutions participating in the {\it Gaia} Multilateral Agreement.

\newpage
\bibliography{ms}{}
\bibliographystyle{aasjournal}

\end{document}